\documentclass[a4paper,11pt]{article}
\usepackage{jheppub}
\usepackage[T1]{fontenc}
\usepackage{tikz,xcolor}
\usepackage{graphicx}
\usepackage{dcolumn}
\usepackage{bm}
\usepackage{hyperref}
\usepackage[title]{appendix}
\usepackage{physics}
\hypersetup{
    colorlinks=true,
    urlcolor= blue,
    citecolor=blue,
    linkcolor= blue}
\def\d{\mbox{\rm d}}
\def\e{\mbox{\rm e}}
\NewDocumentCommand{\tens}{t_}
{%
	\IfBooleanTF{#1}
	{\tensop}
	{\otimes}%
}
\NewDocumentCommand{\tensop}{m}
{%
	\mathbin{\mathop{\otimes}\displaylimits_{#1}}%
}

\definecolor{lime}{HTML}{A6CE39}
\DeclareRobustCommand{\orcidicon}{%
	\begin{tikzpicture}
	\draw[lime, fill=lime] (0,0) 
	circle [radius=0.16] 
	node[white] {{\fontfamily{qag}\selectfont \tiny ID}};
	\draw[white, fill=white] (-0.0625,0.095) 
	circle [radius=0.007];
	\end{tikzpicture}
	\hspace{-2mm}
}

\foreach \x in {A, ..., Z}{%
	\expandafter\xdef\csname orcid\x\endcsname{\noexpand\href{https://orcid.org/\csname orcidauthor\x\endcsname}{\noexpand\orcidicon}}
}

\title{\textbf{\boldmath Transition rates and their applications in accelerated single-qubit for fermionic spinor field coupling}}
\author[a,*]{Arnab Mukherjee\orcidA{}\note[*]{Corresponding author.},}
\author[a]{Sunandan Gangopadhyay\orcidB{},}
\author[b]{P. H. M. Barros\orcidC{},}
\author[b]{and H.~A.~S.~Costa\orcidD{}}
\affiliation[a]{\textit{ S.~N.~Bose National Centre for Basic Sciences, JD Block, Sector-III, Salt Lake, Kolkata 700106, India}}
\affiliation[b]{Departamento de F\'{i}sica, Universidade Federal do Piau\'{i} (UFPI), Campus Min. Petr\^{o}nio Portella - Ininga, Teresina - PI, 64049-550 - Brazil}
\emailAdd{arnab.mukherjee@bose.res.in}
\emailAdd{sunandan.gangopadhyay@bose.res.in}
\emailAdd{phmbarros@ufpi.edu.br}
\emailAdd{hascosta@ufpi.edu.br}
\abstract{\noindent In this work, we investigate the interaction between a uniformly accelerated single qubit and a fermionic spinor field. Here we consider both the massless and the massive fermionic spinor fields. The qubit-field interaction occurs over a finite time and was evolved via perturbation theory. This approach yields the transition probability rates, from which we subsequently evaluate the quantum coherence of an Unruh–DeWitt (UDW) detector initially prepared in a qubit state. Our findings reveal that the UDW detector responds more when coupled with the fermionic field, and consequently, quantum coherence (for the fermionic case) degrades much more rapidly when compared to the case of the qubit linearly coupled with the scalar field. Moreover, the analysis suggests that particle mass plays a protective role against Unruh-induced decoherence as the rest mass energy becomes comparable to the detector’s energy-level spacing, the detector’s excitation probability and response decreases, which leads to the mitigation of quantum coherence degradation in accelerated quantum systems.}

\begin{document}
    \maketitle
    \flushbottom
\section{\label{sec:level1}Introduction}
\noindent Modern theoretical physics continues to reveal profoundly counterintuitive phenomena emerging at the interface between quantum mechanics and general relativity. One of the most striking examples is the so-called Fulling–Davies–Unruh effect~\cite{fulling1973, Davies_1975, unruh1976}, which asserts that a uniformly accelerated observer in the quantum vacuum perceives thermal radiation, as if immersed in a bath of particles, an effect absent for inertial observers~\cite{Crispino2008}. This prediction, first formulated by William Unruh in 1976~\cite{unruh1976}, establishes that particle detection is intrinsically linked to the presence of a Rindler horizon in the accelerated reference frame, a structure that is mathematically analogous to the event horizon of black holes~\cite{birrell1984quantum}. Consequently, distinct quantization procedures lead to ambiguous definitions of particle number~\cite{grove1983notes, wald1994quantum, lancaster2014quantum}.

To formally investigate such effects and other quantum phenomena arising in quantum field theory within non-inertial reference frames, the Unruh–DeWitt (UDW) detector model was introduced~\cite{unruh1976, DeWitt1980}. It provides a simplified framework in which a two-level quantum system interacts with the surrounding quantum fields.
Despite its simplicity, this model encapsulates the essential physics of how different observers interact with quantum fields. The UDW detector functions as a conceptual probe for identifying and quantifying the particles perceived as a consequence of acceleration. In this sense, particles can be operationally defined as those entities detected by a particle detector~\cite{davies1984}.

In recent years, relativistic quantum information (RQI) has emerged as a novel research field that bridges gravitational physics and quantum information. It aims to investigate the influence of relativistic effects on quantum information processing protocols~\cite{martin2011relativistic, Mann2012}. Even without invoking full quantum gravity frameworks, RQI encompasses a broad spectrum of studies with diverse objectives, including the use of quantum probes to analyze the Unruh and Hawking effects~\cite{hawking1975particle, unruh1976, DeWitt1980, wald1994quantum}, as well as the formulation of relativistic quantum communication and computation protocols~\cite{Adlam2015, Landulfo2016, Martinez2015, Martinez2016, Martinez2020, Martinez2021, Tjoa2022, Lapponi2023}. Furthermore, Unruh’s seminal work~\cite{Unruh1981Experimental} drew attention to the thermal emission expected from black holes, demonstrating that analogous systems (such as sonic horizons in fluids) also exhibit comparable thermal radiation. Recently, extensive investigations of the FDU effect under various system configurations have been carried out in Refs.~\cite{mukherjee2023fulling, mukherjee2025single}.

In recent years, quantum coherence has been formally recognized as a resource in its own right, with broad applications in quantum thermodynamics, condensed matter physics, biological systems, and quantum computation~\cite{bibak2024quantum, shi2022entanglement, ahnefeld2022coherence, karli2024controlling, yamauchi2024modulation, wang2024physical, mukherjee2024quantum, mukherjee2025quantum}. Quantum coherence, broadly speaking, originates from the superposition of basis states and serves as a key signature of quantum nature~\cite{leggett1980macroscopic}. A variety of methodologies have been proposed for detecting particles associated with the Unruh effect. Among these, the quantum coherence framework has received increasing attention~\cite{tian2012unruh, Wang2016, he2018multipartite, Nesterov2020, Zhang2022, huang2022, Harikrishnan2022, xu2023decoherence, Pedro2024robustness, barros2024dispersive, barros2024detecting, pedro2025mitigating, barros2025quadratic, wu2025can, liu2025does, li2025multiqubit, li2024bosonic, wu2024does, li2025entropic, barros2025velocity}, in this context, the qubit state is widely used in most of these studies. Within this approach, a UDW detector interacts with the quantum field through particle absorption, thereby modifying its eigenstates and inducing a loss of quantum coherence. Recently, it has also been observed that the degradation of quantum coherence significantly influences the work output of relativistic quantum Otto engines~\cite{mukherjee2022unruh} and the overall performance of relativistic quantum batteries~\cite{mukherjee2024enhancement}.

In general, accelerated detectors are usually studied considering a linear coupling of the detector with a scalar field. In a small number of studies \cite{hummer2016renormalized, sachs2017entanglement, gray2018scalar, wu2023birth, mukherjee2024enhancement, barros2025quadratic}, quadratic coupling interaction between the detector and the quantum fields have been considered. Along these lines, what if we consider a detector that interacts with a fermionic spinorial field? And going even further, what if that field is massive? Regarding these questions, several studies have been conducted, such as: Rindler noise of Dirac Field~\cite{takagi1985response, Takagi1986vacuum}, fermonic and bosonic fields linearly and quadratically coupled \cite{hummer2016renormalized} (presenting a renormalization method), scalar and fermionic Unruh Otto engines~\cite{gray2018scalar}, radiation from a receding mirror with a Dirac fermion~\cite{mokhtar2020radiation}, UDW detectors coupled with a spinor field in (3+1)-dimensional spacetime~\cite{wu2023accelerating}, among others. In summary, these studies found that the detector coupled with a fermionic spinorial field responds differently when compared to the case of linear coupling. This occurs because the fermionic field is a vector field and obeys different statistics than the case of a scalar field. On the other hand, similar effects have also been observed recently in the context of studying black holes and the fermionic spinorial field~\cite{araujo2025non, ARAUJOFILHO2025, araujo2025does, araujo2025particle, ARAUJOFILHO2025117174, Heidari:2025iiv, AraujoFilho:2025jcu}.

\noindent Motivated by these studies, in this work we investigate the effects of coupling a fermionic spinorial field (massless and massive) on the response of a UDW detector and on the quantum coherence of an accelerated single-qubit. 

The paper is structured as follows: In Sec.~\ref{sec:level2}, we analytically calculate the transition probability rates for fermionic coupling, obtaining for both massless and massive coupling cases. In Sec.~\ref{sec:level3}, we use the transition probability rates and apply them to the configuration of an accelerated single-qubit; here we analytically obtain the reduced density matrix of the system and calculate the quantum coherence of the accelerated qubit state. Subsequently, in Sec.\ref{sec:level4}, we use the final analytical expressions of our findings and obtain numerical results through plots, and we discuss these results for all the cases studied. Finally, the last section contains the conclusions of our work.

For convenience, we adopt natural units throughout this work, setting $c = \hbar = k_B = 1$, and use the metric signature $\eta^{\mu\nu} = (+,-,-,-)$.

\section{\label{sec:level2}Transition probability rates with fermionic coupling}
\noindent Initially, we begin by considering an UDW detector, which is modeled as a point-like detector with two energy levels, namely, the ground state $\vert g\rangle$ and the excited state $\vert e\rangle$. In this framework, $\tau$ denotes the proper time of the detector as it moves along a worldline specified by $x(\tau)$. We consider the case in which the detector interacts with a fermionic spinor field $\Psi[x(\tau)]$ through its monopole moment $\mu(\tau)$. Then the Hamiltonian describing the interaction is given by
\begin{eqnarray}
  \mathcal{H}_{\mathrm{int}} = \lambda_{\Psi}\chi(\tau)\mu(\tau) :\overline{\Psi}[x(\tau)] \Psi[x(\tau)]:~,
  \label{Hint}
\end{eqnarray}
where $\lambda_{\Psi}$ is the coupling strength, $\overline{\Psi} = \Psi^{\dagger}\gamma^{0}$ with $\gamma^{\mu}$ being the Dirac matrices that satisfy $\{\gamma^{\mu}, \gamma^{\nu}\} = -2\eta^{\mu\nu}$, and $\chi(\tau)$ is the window function that turns the detector on and off. Due to the finite duration of the interaction, this function satisfies the following conditions: $\chi(\tau) \approx 1$ for $|\tau| \ll T$, and $\chi(\tau) \approx 0$ for $|\tau| \gg T$, where $T$ is the finite interaction time. In Eq.~(\ref{Hint}), there is the standard field theoretic method of normal ordering~\cite{hummer2016renormalized}, defined by~$:\hat{A}: = \hat{A} - \langle 0_{\mathcal{M}} \vert \hat{A} \vert 0_{\mathcal{M}} \rangle$ for an operator $\hat{A}$, with $\vert 0_{\mathcal{M}} \rangle$ being the Minkowski vacuum. The application of normal ordering is essential, as persistent divergences arise in the context of fermionic detector couplings, which cannot be regularized by means of either the switching function or spatial smearing of the detector~\cite{hummer2016renormalized, Takagi1986vacuum, Louko2016}.

Additionally, as noted by \cite{gray2018scalar}, each type of coupling leads to a different dimensionality, i.e, the coupling constants also possess distinct dimensions. Specifically, for dimensionless switching functions, it can be shown that $[\lambda_{\Psi}] \equiv -\Delta_{\Psi} = 2 - d$, where $d$ denotes the spacetime dimension, and we adopt units such that $[x] = 1$. Nevertheless, it is possible to define a dimensionless coupling parameter of the form $\overline{\lambda}_{\Psi} = \lambda_{\Psi}/\zeta^{\Delta_{\Psi}}$, where $\zeta$ represents a characteristic time scale. In this work, we take $d = 4$, and the relevant time scale for detector evolution is $\zeta \sim \Omega^{-1}$. Consequently, we obtain the dimensionless coupling $\overline{\lambda}_{\Psi} = \lambda_{\Psi} \Omega^2$. Note that adopting $\Omega = 1$ the constant $\overline{\lambda}_{\Psi}$ for fermionic coupling is equal to the constant $\lambda_{\phi}$ of scalar coupling.

\noindent In this way, considering an interaction up to first order of the detector with a fermionic spinor field, we can obtain the probabilities of excitation (absorption) $\mathcal{P^{-}}$ and de-excitation (emission) $\mathcal{P^{+}}$ given by
\begin{eqnarray}
  \mathcal{P}^{\pm}_{\Psi} = \lambda^2_{\Psi}|\langle g|\mu(0)|e\rangle|^2 \mathcal{F}^{\pm}_{\Psi}~,
\label{P}
\end{eqnarray}
where
\begin{eqnarray}
    \mathcal{F}^{\pm}_{\Psi} = \int^{+\infty}_{-\infty}\d\tau \int^{+\infty}_{-\infty}\d\tau' \chi(\tau)\chi(\tau')\e^{\pm i\Omega(\tau-\tau')} \mathcal{W}_{\Psi}(x,x')~,~
    \label{Response_Function}
\end{eqnarray}
with $\Omega$ being the transition angular frequency, with excitation processes corresponding to $\Omega > 0$ and de-excitation processes to $\Omega < 0$ also $x\equiv x(\tau)$ and $x'\equiv x(\tau')$.
\subsection{Coupling with massless fermionic field}
\noindent In this subsection, we will discuss the scenario where the UDW detector is coupled with the massless fermionic spinor fields. 

For the massless fermionic coupling, the Wightman function \cite{Wightman1956} is given by,
\begin{eqnarray}
    \mathcal{W}_{\Psi,\,0}(x,x') = \langle 0_{\mathcal{M}}\vert :\overline{\Psi}\Psi[x(\tau)]::\overline{\Psi}\Psi[x(\tau')]: \vert 0_{\mathcal{M}}\rangle~,
    \label{defWightman}
\end{eqnarray}
and following ref.~\cite{Louko2016, gray2018scalar}, we have
\begin{equation}
    \mathcal{W}_{\Psi,\,0}(x,x') = \mathrm{Tr}[S^{+}(x,x')\,S^{-}(x',x)]~\label{e_5}
\end{equation}
and using the conventions of \cite{Takagi1986vacuum}, the Dirac field Wightman functions $S^{+}(x,x')$ and $S^{-}(x',x)$ are given by
\begin{eqnarray}
    S^{+}_{0,\,ab}(x,x') &=& +(i\gamma^{\mu}\partial_{x^\mu})_{ab}\mathcal{W}_{\phi,\,0}(x,x')~,~\label{e_6} \\[6pt]
    S^{-}_{0,\,ab}(x',x) &=& -(i\gamma^{\mu}\partial_{x'^\mu})_{ab}\mathcal{W}_{\phi,\,0}(x,x')~,~\label{e_7}
\end{eqnarray}
where $\mathcal{W}_{\phi,\,0}(x,x')$ is the (well-known) Wightman function of a real and massless scalar field. 

\noindent Using results given in eqs. \eqref{e_6}, and \eqref{e_7} into eq.~\eqref{e_5}, we obtain
\begin{equation}
    \mathcal{W}_{\Psi,\,0}(x,x') =4\eta^{\mu\nu}\partial_{x^\mu}\mathcal{W}_{\phi,\,0}(x,x')\,\partial_{x'^\nu}\mathcal{W}_{\phi,\,0}(x,x')~.~\label{e_8}
\end{equation}
In case of the real massless scalar field, the Wightman function is given by \cite{birrell1984quantum}
\begin{equation}
    \mathcal{W}_{\phi,\,0}(x,x')=-\frac{1}{4\pi^2}\left[\frac{1}{(t-t'-i\epsilon)^2-|\vec{x}-\vec{x}'|^2}\right]~.~\label{e_9}
\end{equation}
Using eq.~\eqref{e_9} in eq.~\eqref{e_8}, and taking the derivatives properly, the Wightman function for the massless fermionic field takes the form
\begin{equation}
    \mathcal{W}_{\Psi,\,0}(x,x') =-\frac{1}{\pi^4}\left[\frac{1}{(t-t'-i\epsilon)^2-|\vec{x}-\vec{x}'|^2}\right]^{3}~.~\label{e_10}
\end{equation}
The general result of the Wightman function for the massless fermionic field in $d$ dimension is given by \cite{gray2018scalar}
\begin{eqnarray}
    \mathcal{W}_{\Psi,\,0}(x,x') = \frac{N_d \Gamma(d/2)^2}{\Gamma(d-1)}\mathcal{W}^{2d}_{\phi}(x,x'),
    \label{e_11}
\end{eqnarray}
with $N_d = 2^{d/2}$ in even ($N_d = 2^{(d-1)/2}$ in odd) dimensions and $\mathcal{W}^{2d}_{\phi,\,0}(x,y)$ is $\mathcal{W}_{\phi,\,0}(x,y)$ in $2d$ dimensions. If the worldline along which the detector moves corresponds to an integral curve of a timelike Killing vector field, then eq.~(\ref{defWightman}) becomes invariant under time translations in the detector’s frame of reference and depends solely on the proper time difference \cite{letaw1981quantized, padmanabhan1982general}, i.e., $\mathcal{W}_{\Psi}[x(\tau), x(\tau')] = \mathcal{W}_{\Psi}(\tau, \tau') = \mathcal{W}_{\Psi}(\Delta\tau)$, with $\Delta\tau = \tau - \tau'$.

However, when applying these results to a trajectory undergoing constant and uniform acceleration $a$ in the $z$-direction, the intrinsic coordinate system of the detector's frame is related to the Minkowski coordinates via the Rindler transformations \cite{rindler1966kruskal}, given by
\begin{eqnarray}
    t(\tau) = \frac{1}{a} \sinh{a\tau}, \quad x(\tau) = \frac{1}{a} \cosh{a\tau}, \quad y(\tau) = z(\tau) = 0~.~
    \label{e_12}
\end{eqnarray}
Now, using the above trajectory eq.~\eqref{e_12} in eq.~\eqref{e_10}, the Wightman function for the massless fermionic field takes the form
\begin{eqnarray}
    \mathcal{W}_{\Psi,\,0}(\Delta\tau) = -\frac{a^{6}}{64\pi^{4}}\sinh^{-6}\left( \frac{a\Delta\tau-i\epsilon}{2}\right)~.\label{e_13}
\end{eqnarray}
The series representation of $\sinh^{-2}\left(\frac{a\Delta\tau-i\epsilon}{2}\right)$ can be written as \cite{gradshteyn2014table} 
\begin{equation}
    \frac{1}{\sinh^{2}\left(\frac{a\Delta\tau-i\epsilon}{2}\right)}=4\sum^{\infty}_{k=-\infty}\frac{1}{(a\Delta\tau-i\epsilon-2i\pi k)^2}~.\label{e_14}
\end{equation}
Using the above series representation, following relations can be obtained
\begin{align}
    \frac{1}{\sinh^{4}\left(\frac{a\Delta\tau-i\epsilon}{2}\right)}&=\,16\sum^{\infty}_{k=-\infty}\left[\frac{1}{(a\Delta\tau-i\epsilon-2i\pi k)^4}-\frac{1}{6}\,\frac{1}{(a\Delta\tau-i\epsilon-2i\pi k)^2}\right]~,\label{e_15}\\[6pt]
    \frac{1}{\sinh^{6}\left(\frac{a\Delta\tau-i\epsilon}{2}\right)}&=\,64\sum^{\infty}_{k=-\infty}\left[\frac{1}{(a\Delta\tau-i\epsilon-2i\pi k)^6}-\frac{1}{4}\,\frac{1}{(a\Delta\tau-i\epsilon-2i\pi k)^4}\right.\nonumber\\
    &+\left.\frac{1}{30}\,\frac{1}{(a\Delta\tau-i\epsilon-2i\pi k)^2}\right]~.~\label{e_16}
\end{align}
Inserting eq.~\eqref{e_16} into eq.~\eqref{e_13}, finally we have the following expression
\begin{align}
    \mathcal{W}_{\Psi,\,0}(\Delta\tau) 
    &=-\frac{a^6}{\pi^{4}}\sum^{\infty}_{k=-\infty} \Bigg[ \frac{1}{(a\Delta\tau-i\epsilon-2i\pi k)^6}- \frac{1}{4} \frac{1}{(a\Delta\tau-i\epsilon-2i\pi k)^4}\nonumber\\
    &+\frac{1}{30} \frac{1}{(a\Delta\tau-i\epsilon-2i\pi k)^2}\Bigg]~.
    \label{e_17}
\end{align}
This expression is the Wightman function for the case of a massless fermionic spinor field. Take note that it satisfies the Kubo-Martin-Schwinger (KMS) condition \cite{kubo1957statistical, Martin1959Schwinger}.

\noindent By employing power series techniques in eq.~(\ref{Response_Function}) and expanding $\chi(\tau)$ in a Taylor series around $\tau = 0$, the transition probabilities can be expressed as \cite{padmanabhan1982general, sriramkumar1996finite}
\begin{eqnarray}
    \mathcal{F}^{\pm}_{\Psi} \approx \mathcal{F}^{\pm}_{\Psi}(\infty) - \chi''(0)\frac{\partial^2\mathcal{F}^{\pm}_{\Psi}(\infty)}{\partial\Omega^2}~,
\end{eqnarray}
where $\mathcal{F}^{\pm}_{\Psi}(\infty)$ is the detector response function for infinite time, and their corresponding transition probability per unit time is given by
\begin{eqnarray}
\mathcal{R}_{\Psi}^{\pm} \approx \mathcal{R}_{\Psi}^{\pm}(\infty) - \chi''(0)\frac{\partial^2 \mathcal{R}_{\Psi}^{\pm}(\infty)}{\partial\Omega^2}~,\label{e_19}
\end{eqnarray}
where
\begin{eqnarray}
    \mathcal{R}_{\Psi}^{\pm}(\infty) = \lambda^2_{\Psi}\int_{-\infty}^{\infty} \d(\Delta\tau) \e^{\pm i\Omega\Delta\tau}\mathcal{W}_{\Psi}(\Delta\tau)~.\label{e_20}
\end{eqnarray}
An abrupt activation or deactivation of the detector may lead to divergent contributions. To circumvent such divergences, we adopt a Gaussian switching function defined by $\chi(\tau) = \exp\left(-\frac{\tau^2}{2T^2}\right)$, which ensures a smooth turn-on and turn-off of the detector. With this choice, and to leading order, the finite-time corrections to the transition probability rate can be expressed as
\begin{eqnarray}
 \mathcal{R}_{\Psi}^{\pm} \approx \mathcal{R}_{\Psi}^{\pm}(\infty) + \frac{1}{2T^2}\frac{\partial^2 \mathcal{R}_{\Psi}^{\pm}(\infty)}{\partial\Omega^2} + \mathcal{O}\left(T^{-4}\right)~,
 \label{e_21}
\end{eqnarray}
where $\mathcal{R}_{\Psi}^{-}$ and $\mathcal{R}_{\Psi}^{+}$ being the rates of the excitation probability and the de-excitation probability, respectively.\\[2pt] 
Using eq.~\eqref{e_17} in eq.~(\ref{e_20}), we have
\begin{align}
    \mathcal{R}_{\Psi,\,0}^{-}(\infty) &= \frac{1}{60\pi^3}\left( 1 + \frac{5a^2}{\Omega^2} + \frac{4a^4}{\Omega^4}\right)\frac{\Omega^5}{\e^{2\pi\Omega/a}-1}~, \label{e_22}\\[6pt]
    \mathcal{R}_{\Psi,\,0}^{+}(\infty) &= \frac{\e^{2\pi\Omega/a}}{60\pi^3}\left( 1 + \frac{5a^2}{\Omega^2} + \frac{4a^4}{\Omega^4}\right)\frac{\Omega^5}{\e^{2\pi\Omega/a}-1}~,\label{e_23}
\end{align}
and, in this way, take note that $\mathcal{R}_{\Psi}^{\pm} \sim \Omega^5$ and
\begin{equation}
    \mathcal{R}_{\Psi,\,0}^{\pm}(\infty)= \frac{\Omega^4}{30\pi^2}\left( 1 + \frac{5a^2}{\Omega^2} + \frac{4a^4}{\Omega^4}\right)\mathcal{R}_{\phi}^{\pm}~,\label{e_24}
\end{equation}
where $\mathcal{R}_{\phi,\,0}^{\pm}$ is the transition probability rates for scalar field coupling.\\[2pt]
Substituting eq.~\eqref{e_22} and \eqref{e_23} into eq.~\eqref{e_21}, and after long and tedious algebraic manipulations, we obtain
\begin{align}
    \mathcal{R}^{-}_{\Psi,\,0} &\approx \frac{\Omega^{5}}{60\pi^{3}}\frac{(1 + 5\overline{a}^2 + 4\overline{a}^4)}{(\e^{2\pi/\overline{a}} - 1)} \Bigg[1 + \frac{10(1 + \frac{3}{2}\overline{a}^2)}{\sigma^2(1 + 5\overline{a}^2 + 4\overline{a}^4)}-\frac{10\pi}{\overline{a}\sigma^2} \frac{\e^{2\pi/\overline{a}}}{(\e^{2\pi/\overline{a}} - 1)} \frac{(1 + 3\overline{a}^2 + \frac{4}{5}\overline{a}^4)}{(1 + 5\overline{a}^2 + 4\overline{a}^4)} \nonumber\\
    &\times\Bigg\{ 1 - 
\frac{\pi}{5\overline{a}} \left( \frac{\e^{2\pi/\overline{a}}+1}{\e^{2\pi/\overline{a}} -1}\right) \frac{(1 + 5\overline{a}^2 + 4\overline{a}^4)}{(1 + 3\overline{a}^2 + \frac{4}{5}\overline{a}^4)} \Bigg\}\Bigg]~,~\label{e_25}
\end{align}
and similarly,
\begin{align}
    \mathcal{R}^{+}_{\Psi,\,0} &\approx \frac{\Omega^{5} \e^{2\pi/\overline{a}}}{60\pi^{3}}\frac{(1 + 5\overline{a}^2 + 4\overline{a}^4)}{(\e^{2\pi/\overline{a}} - 1)} \Bigg[1 + \frac{10(1 + \frac{3}{2}\overline{a}^2)}{\sigma^2(1 + 5\overline{a}^2 + 4\overline{a}^4)} -\frac{10\pi}{\overline{a}\sigma^2} \frac{\e^{2\pi/\overline{a}}}{(\e^{2\pi/\overline{a}} - 1)} \frac{(1 + 3\overline{a}^2 + \frac{4}{5}\overline{a}^4)}{(1 + 5\overline{a}^2 + 4\overline{a}^4)} \nonumber\\
    &\times\Bigg\{ 1 - 
\frac{\pi}{5\overline{a}} \left( \frac{\e^{2\pi/\overline{a}}+1}{\e^{2\pi/\overline{a}} -1}\right) \frac{(1 + 5\overline{a}^2 + 4\overline{a}^4)}{(1 + 3\overline{a}^2 + \frac{4}{5}\overline{a}^4)} \Bigg\}\Bigg]~,~\label{e_26}
\end{align}
where we define the following dimensionless parameters,  $\overline{a} \equiv a/\Omega$ and $\sigma \equiv \Omega T$.\\[2pt] 
The expressions in eqs.~\eqref{e_25} and \eqref{e_26} characterize the transition probability rates between the internal states of a detector that interacts with the fermionic spinor field over a finite duration.

It is important to emphasize that, for a finite interaction time, the detector does not have sufficient time to reach complete thermal equilibrium with the quantum field. As a result, the response deviates slightly from that expected for ideal thermal radiation \cite{Svaiter1992Inertial, Higuchi1993Uniformly, padmanabhan1982general}. For a comparison between the linear and quadratic cases of transition probability rates, see the corresponding results for the linear case in \cite{padmanabhan1982general, Pedro2024robustness}.
\subsection{Coupling with massive fermionic field}
\noindent In this subsection, we investigate the interaction between a UDW detector and massive fermionic spinor fields.
For the massive fermionic coupling, $S^{+}(x,x')$ and $S^{-}(x',x)$ are given by
\begin{align}
    S^{+}_{m,\,ab}(x,x') &= +(i\gamma^{\mu}\partial_{x^\mu}+m)_{ab}\mathcal{W}_{\phi,\,m}(x,x')~,~\label{e_27} \\[6pt]
    S^{-}_{m,\,ab}(x',x) &= -(i\gamma^{\mu}\partial_{x'^\mu}+m)_{ab}\mathcal{W}_{\phi,\,m}(x,x')~,~\label{e_28}
\end{align}
where $\mathcal{W}_{\phi,\,m}(x,x')$ is the Wightman function of a real and massive scalar field.\\[2pt]
Using results given in eqs. \eqref{e_27}, and \eqref{e_28} into eq.~\eqref{e_5}, we obtain
\begin{equation}
    \mathcal{W}_{\Psi,\,m}(x,x') =4\eta^{\mu\nu}\partial_{x^\mu}\mathcal{W}_{\phi,\,m}(x,x')\,\partial_{x'^\nu}\mathcal{W}_{\phi,\,m}(x,x')-4m^2(\mathcal{W}_{\phi,\,m}(x,x'))^2~.\label{e_29}
\end{equation}
In case of the massive scalar field, the Wightman function is given by \cite{birrell1984quantum, alkofer2016quantum}
\begin{equation}
    \mathcal{W}_{\phi,\,m}(x,x')=\frac{m}{4\pi^2}\left[\frac{\mathcal{K}_{1}\left(m\sqrt{-\rho}\right)}{\sqrt{-\rho}}\right]~,~\label{e_30}
\end{equation}
where $\rho=(t-t'-i\epsilon)^2-|\vec{x}-\vec{x}'|^2$, and $\mathcal{K}_1$ is the modified Bessel function of the second kind of first order. At this point, it is noted that in the massless limit, eq.~\eqref{e_30} reduces to the Wightman function of the massless scalar field given in eq.~\eqref{e_9}. See Appendix~\ref{App:A} for the derivation of eq.~\eqref{e_30}.

Using eq.~\eqref{e_30} in eq.~\eqref{e_29}, and taking the derivatives properly, the Wightman function for the massive fermionic field takes the form
\begin{equation}
    \mathcal{W}_{\Psi,\,m}(x,x')=\frac{m}{4\pi^{4}\rho}\bigg[\mathcal{K}^{2}_{1}\left(m\sqrt{-\rho}\right)-\mathcal{K}^{2}_{2}\left(m\sqrt{-\rho}\right)\bigg]~.\label{e_31}
\end{equation}
Here, $\mathcal{K}_2$ is the modified Bessel function of the second kind of second order. See Appendix~\ref{App:B} for the derivation of eq.~\eqref{e_31}.
Inserting the trajectory of the uniformly accelerated detector given in eq.~\eqref{e_12} into eq.~\eqref{e_31}, we obtain
\begin{align}
    \mathcal{W}_{\Psi,\,m}(\Delta\tau)&=\frac{m^{4}a^{2}}{16\pi^{4}}\sinh^{-2}\left( \frac{a\Delta\tau-i\epsilon}{2}\right)\bigg[\mathcal{K}^{2}_{1}\left(i\frac{2m}{a}\sinh\left( \frac{a\Delta\tau-i\epsilon}{2}\right)\right)\nonumber\\
    &-\mathcal{K}^{2}_{2}\left(i\frac{2m}{a}\sinh\left( \frac{a\Delta\tau-i\epsilon}{2}\right)\right)\bigg]~.\label{e_32}
\end{align}
In order to compute the excitation and de-excitation rates by using eq.~\eqref{e_32} into eqs.~\eqref{e_20} and \eqref{e_21}, at first we approximate eq.~\eqref{e_32} into \textit{small-mass} $(\frac{m}{a}\ll1)$ and \textit{large-mass} $(\frac{m}{a}\gg1)$ limits.

For small argument limit, modified Bessel function of the second kind $\mathcal{K}_{\nu}(z)$ can be written up to the leading order as \cite{abramowitz1964handbook}
\begin{equation}
    \mathcal{K}_{\nu}(z)\sim \frac{1}{2}\,\Gamma(\nu)\,\left(\frac{z}{2}\right)^{-\nu}~\label{e_33}
\end{equation}
where $\nu$ has fixed values and $\nu>0$. Similarly, for large argument limit, up to the leading order modified Bessel function of the second kind $\mathcal{K}_{\nu}(z)$ can be written as \cite{abramowitz1964handbook}
\begin{equation}
    \mathcal{K}_{\nu}(z)\sim \sqrt{\frac{\pi}{2z}}\,e^{-z}\left(1+\frac{4\nu^{2}-1}{8z}+\mathcal{O}\left(\frac{1}{z^2}\right)\right)~.\label{e_34}
\end{equation}
\subsubsection{Small-mass approximation}
\noindent Using the small argument expansion of the 
modified Bessel function of the second kind $\mathcal{K}_{\nu}(z)$ (eq.~\eqref{e_33}) in eq.~\eqref{e_32}, we obtain
\begin{equation}
    \mathcal{W}_{\Psi,\,sm}(\Delta\tau)\approx-\frac{a^{6}}{64\pi^{4}}\sinh^{-6}\left( \frac{a\Delta\tau-i\epsilon}{2}\right)-\frac{m^{2}a^{4}}{64\pi^{4}}\sinh^{-4}\left( \frac{a\Delta\tau-i\epsilon}{2}\right)+\mathcal{O}(m^4)~.\label{e_35}
\end{equation}
Here, ``\textit{sm}'' subscript stands for ``\textit{small-mass}'' limit. Using the series representations given in eqs.~\eqref{e_15}, \eqref{e_16} into eq.~\eqref{e_35}, we obtain Wightman function up to $\mathcal{O}(m^2)$ as
\begin{align}
    \mathcal{W}_{\Psi,\,sm}(\Delta\tau) 
    &=-\frac{a^6}{\pi^{4}}\sum^{\infty}_{k=-\infty} \Bigg[ \frac{1}{(a\Delta\tau-i\epsilon-2i\pi k)^6}- \frac{1}{4}\left(1-\frac{m^2}{a^2}\right)\frac{1}{(a\Delta\tau-i\epsilon-2i\pi k)^4}~\nonumber\\
    &+\frac{1}{30}\left(1-\frac{5m^2}{4a^2}\right)\frac{1}{(a\Delta\tau-i\epsilon-2i\pi k)^2}\Bigg]~.
    \label{e_36}
\end{align}
Using eq.~\eqref{e_36} in eq.~(\ref{e_20}), we have
\begin{align}
    \mathcal{R}_{\Psi,\,sm}^{-}(\infty) &= \frac{1}{60\pi^3}\bigg[\left(1-\frac{5m^2}{\Omega^2}\right) +5\left(1-\frac{m^2}{\Omega^2}\right)\frac{a^2}{\Omega^2} + \frac{4a^4}{\Omega^4}\bigg]\left(\frac{\Omega^5}{\e^{2\pi\Omega/a}-1}\right)~, \label{e_37}\\[6pt]
    \mathcal{R}_{\Psi,\,sm}^{\,+}(\infty) &= \frac{\e^{2\pi\Omega/a}}{60\pi^3}\bigg[\left(1-\frac{5m^2}{\Omega^2}\right) +5\left(1-\frac{m^2}{\Omega^2}\right)\frac{a^2}{\Omega^2} + \frac{4a^4}{\Omega^4}\bigg]\left(\frac{\Omega^5}{\e^{2\pi\Omega/a}-1}\right)~.\label{e_38}
\end{align}
Substituting eq.~\eqref{e_37} into eq.~\eqref{e_21}, we obtain the excitation rate for massive fermionic coupling in the small-mass limit in the form
\begin{align}
\mathcal{R}_{\Psi,\,sm}^{\,-}&\approx \mathcal{R}_{\Psi,\,0}^{-}-\frac{\overline{m}^2\Omega^{5}}{12\pi^{3}}\frac{(1 + \overline{a}^2)}{(\e^{2\pi/\overline{a}} - 1)} \Bigg[1 + \frac{3}{\sigma^2(1+\overline{a}^2)}-\frac{2\pi}{\overline{a}\sigma^2}\frac{(3+\overline{a}^2)}{(1+\overline{a}^2)} \frac{\e^{2\pi/\overline{a}}}{(\e^{2\pi/\overline{a}}-1)}  \nonumber\\
    &\times\Bigg\{1-\frac{\pi}{\overline{a}} \left( \frac{\e^{2\pi/\overline{a}}+1}{\e^{2\pi/\overline{a}} -1}\right) \frac{(1+\overline{a}^2)}{(3+\overline{a}^2)}\Bigg\}\Bigg]~,~\label{e_39}
\end{align}
where $\mathcal{R}_{\Psi}^{-}$ is given in eq.~\eqref{e_25}. Similarly, de-excitation rate for massive fermionic coupling in the small-mass limit is given by
\begin{align}
\mathcal{R}_{\Psi,\,sm}^{\,+}&\approx \mathcal{R}_{\Psi,\,0}^{+}-\frac{\overline{m}^2\Omega^{5}\e^{2\pi/\overline{a}}}{12\pi^{3}}\frac{(1 + \overline{a}^2)}{(\e^{2\pi/\overline{a}} - 1)} \Bigg[1 + \frac{3}{\sigma^2(1+\overline{a}^2)}-\frac{2\pi}{\overline{a}\sigma^2}\frac{(3+\overline{a}^2)}{(1+\overline{a}^2)} \frac{\e^{2\pi/\overline{a}}}{(\e^{2\pi/\overline{a}}-1)}  \nonumber\\
    &\times\Bigg\{1-\frac{\pi}{\overline{a}} \left( \frac{\e^{2\pi/\overline{a}}+1}{\e^{2\pi/\overline{a}} -1}\right) \frac{(1+\overline{a}^2)}{(3+\overline{a}^2)}\Bigg\}\Bigg]~,~\label{e_40}
\end{align}
where $\mathcal{R}_{\Psi}^{+}$ is given in eq.~\eqref{e_26}.

From the transition rates given in eqs.~\eqref{e_39} and \eqref{e_40}, it is clearly observed that at the limit $\overline{m}\rightarrow0$ both the equations will be equal to those of the massless scenario given in eqs.~\eqref{e_25} and \eqref{e_26}.
\subsubsection{Large-mass approximation}
\noindent Using the large argument expansion of the 
modified Bessel function of the second kind $\mathcal{K}_{\nu}(z)$ (eq.~\eqref{e_34}) in eq.~\eqref{e_32}, up to leading order we get
\begin{equation}
    \mathcal{W}_{\Psi,\,lm}(\Delta\tau)\approx\frac{3m^{2}a^{4}}{128\pi^{3}}\,\e^{-i\left(\frac{4m}{a}\sinh\left( \frac{a\Delta\tau-i\epsilon}{2}\right)\right)}\sinh^{-4}\left( \frac{a\Delta\tau-i\epsilon}{2}\right)~.\label{e_41}
\end{equation}
Here, ``\textit{lm}'' subscript stands for ``\textit{large-mass}'' limit. Using the series representation given in eq.~\eqref{e_15} into eq.~\eqref{e_41}, we obtain Wightman function up to the leading order as
\begin{align}
    \mathcal{W}_{\Psi,\,lm}(\Delta\tau)
    &\approx\frac{3m^{2}a^{4}}{8\pi^{3}}\,\exp\left[-i\left(\frac{4m}{a}\sinh\left( \frac{a\Delta\tau-i\epsilon}{2}\right)\right)\right]\nonumber\\
    &\times\sum^{\infty}_{k=-\infty}\left[\frac{1}{(a\Delta\tau-i\epsilon-2i\pi k)^4}-\frac{1}{6}\,\frac{1}{(a\Delta\tau-i\epsilon-2i\pi k)^2}\right]~.\label{e_42}
\end{align}
Using eq.~\eqref{e_42} in eq.~\eqref{e_20}, we obtain
\begin{align}
    \mathcal{R}_{\Psi,\,lm}^{-}(\infty) &= \frac{m^2}{8\pi^2}\bigg[1+\frac{a^2}{\Omega^2}-\frac{12m}{\Omega}+\frac{48m^2}{\Omega^2}-\frac{64m^3}{\Omega^3} \bigg]\left(\frac{\Omega^3}{\e^{2\pi\Omega/a}-1}\right)~, \label{e_43}\\[6pt]
    \mathcal{R}_{\Psi,\,lm}^{+}(\infty) &= \frac{m^{2}\e^{2\pi\Omega/a}}{8\pi^2}\bigg[1+\frac{a^2}{\Omega^2}-\frac{12m}{\Omega}+\frac{48m^2}{\Omega^2}-\frac{64m^3}{\Omega^3} \bigg]\left(\frac{\Omega^3}{\e^{2\pi\Omega/a}-1}\right)~.\label{e_44}
\end{align}
Substituting eq.~\eqref{e_43} into eq.~\eqref{e_21}, we obtain the excitation rate for massive fermionic coupling in the large-mass limit in the form
\begin{align}
    &\,\,\mathcal{R}_{\Psi,\,lm}^{-}
    \approx\frac{\overline{m}^{2}\Omega^{5}}{8\pi^2}\,\left(1+\overline{a}^{2}-12\,\overline{m}+48\,\overline{m}^2-64\,\overline{m}^3\right)\frac{1}{(\e^{2\pi/\overline{a}} - 1)}\nonumber\\
    &\times\bigg[1+\frac{3(1-4\overline{m})}{\sigma^{2}(1+\overline{a}^{2}-12\,\overline{m}+48\,\overline{m}^2-64\,\overline{m}^3)}-\frac{2\pi}{\overline{a}\sigma^2}\frac{(3+\overline{a}^{2}-24\,\overline{m}+48\,\overline{m}^2)}{(1+\overline{a}^{2}-12\,\overline{m}+48\,\overline{m}^2-64\,\overline{m}^3)}\nonumber\\
    &\times\bigg\{1-\frac{\pi}{\overline{a}}\left( \frac{\e^{2\pi/\overline{a}}+1}{\e^{2\pi/\overline{a}} -1}\right)\frac{(1+\overline{a}^{2}-12\,\overline{m}+48\,\overline{m}^2-64\,\overline{m}^3)}{(3+\overline{a}^{2}-24\,\overline{m}+48\,\overline{m}^2)}\bigg\}\bigg]~.\label{e_45}
\end{align}
Similarly, de-excitation rate for massive fermionic coupling in the large mass limit is given by
\begin{align}
    &\,\,\mathcal{R}_{\Psi,\,lm}^{+}\approx\frac{\overline{m}^{2}\Omega^{5}}{8\pi^2}\,\left(1+\overline{a}^{2}-12\,\overline{m}+48\,\overline{m}^2-64\,\overline{m}^3\right)\frac{\e^{2\pi/\overline{a}}}{(\e^{2\pi/\overline{a}} - 1)}\nonumber\\
    &\times\bigg[1+\frac{3(1-4\overline{m})}{\sigma^{2}(1+\overline{a}^{2}-12\,\overline{m}+48\,\overline{m}^2-64\,\overline{m}^3)}-\frac{2\pi}{\overline{a}\sigma^2}\frac{(3+\overline{a}^{2}-24\,\overline{m}+48\,\overline{m}^2)}{(1+\overline{a}^{2}-12\,\overline{m}+48\,\overline{m}^2-64\,\overline{m}^3)}\nonumber\\
    &\times\bigg\{1-\frac{\pi}{\overline{a}}\left( \frac{\e^{2\pi/\overline{a}}+1}{\e^{2\pi/\overline{a}} -1}\right)\frac{(1+\overline{a}^{2}-12\,\overline{m}+48\,\overline{m}^2-64\,\overline{m}^3)}{(3+\overline{a}^{2}-24\,\overline{m}+48\,\overline{m}^2)}\bigg\}\bigg]~.\label{e_46}
\end{align} 
From the above equations, it is observed that the transition rates of the large mass limits completely depend on the parameter $\overline{m}$. On the other hand, it is important to emphasize that when we have a mass much greater than the acceleration, that is, the large-mass limit ($m/a \gg 1$), we see in eqs. ~(\ref{e_45}) and (\ref{e_46}) that when $\overline{m} \to \infty$ the detector does not respond ($\overline{\mathcal{R}}_{\Psi,\,lm}^{\pm} \to 0$), because the term $-64\overline{m}^3$ quickly suppresses the transition rate. Thus, for future applications and analyses we will only use the small-mass limit.

\section{\label{sec:level3}Accelerated single-qubit coherence}
\noindent We now perform an investigation into the influence of the Unruh effect on the quantum coherence of a uniformly accelerated single-qubit. In particular, our purpose is to elucidate the impact of fermionic spinor field coupling on the fundamental properties of two-level quantum systems. For this purpose, we consider a model in which a detector interacts with a massless and massive fermionic spinor field. Within this framework, the field is initially prepared in the Minkowski vacuum state $\vert 0_{\mathcal{M}} \rangle$, while the detector is initialized in a general qubit state, namely,
\begin{eqnarray}
|\psi_{\mathrm{D}}\rangle = \alpha|g\rangle + \beta|e\rangle,
\end{eqnarray}
where $\alpha = \e^{i\frac{\varphi}{2}}\cos{\frac{\theta}{2}}$, $\beta = \e^{-i\frac{\varphi}{2}}\sin{\frac{\theta}{2}}$ are complex amplitudes, and here, $\theta \in [0,\pi]$ and $\varphi \in [0,2\pi]$ are the polar and azimuthal angles of the Bloch sphere \cite{jazaeri2019review, kasirajan2021quantum}

\begin{figure}
    \centering
    \includegraphics[width=0.6\linewidth]{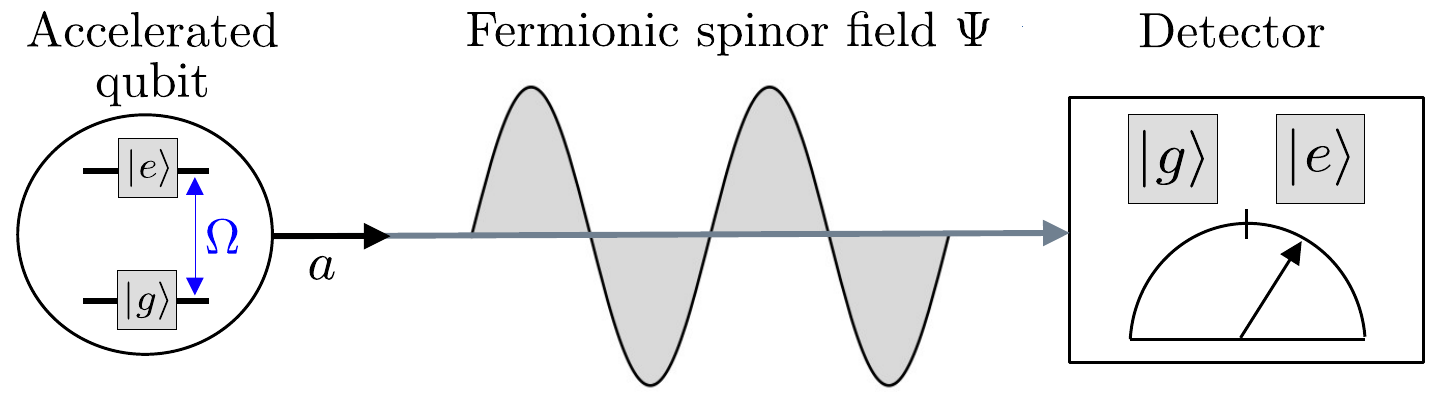}
    \caption{Schematic representation of the measurement of the coherence degradation of an accelerated single-qubit interacting (for a finite time) with a fermionic spinor field.}
    \label{F1}
\end{figure}

As shown by Fig.~\ref{F1}, we consider an UDW detector initially prepared in a general qubit state. This detector interacts with a massless and fermionic spinor field $\Psi$ over a finite time interval $T$. Following the interaction, it becomes possible to measure the detector’s internal states, $\vert g \rangle$ and $\vert e \rangle$. Mathematically, the initial state of the combined system is given by $\hat{\rho}_{\mathrm{in}} = \hat{\rho}^{\mathrm{in}}_{D} \otimes \hat{\rho}_{\Psi}$, where $\hat{\rho}^{\mathrm{in}}_D = |\psi_D\rangle \langle\psi_D|$ and $\hat{\rho}_{\Psi} = |0_{\mathcal{M}} \rangle \langle 0_{\mathcal{M}}|$.

The density matrix after the detector-field interaction is governed by the Hamiltonian that characterizes the fermionic coupling (eq.~(\ref{Hint})), and is given by
\begin{equation}
    \hat{\rho}^{\mathrm{out}}_{\Psi} = \mathcal{\hat{U}}^{(0)}_{\Psi} \hat{\rho}_{\mathrm{in}} \mathcal{\hat{U}}^{(0)^\dagger}_{\Psi} + \mathcal{\hat{U}}^{(1)}_{\Psi} \hat{\rho}_{\mathrm{in}} + \hat{\rho}_{\mathrm{in}}\mathcal{\hat{U}}^{(1)^\dagger}_{\Psi}
    + \mathcal{\hat{U}}^{(1)}_{\Psi} \hat{\rho}_{\mathrm{in}} \mathcal{\hat{U}}^{(1)^\dagger}_{\Psi} + \mathcal{\hat{U}}^{(2)}_{\Psi} \hat{\rho}_{\mathrm{in}} + \hat{\rho}_{\mathrm{in}} \mathcal{\hat{U}}^{(2)^\dagger}_{\phi^2} + \mathcal{O}(\lambda^3_{\Psi})~.
\end{equation}
In this context, $\hat{\mathcal{U}}_{\Psi}$ denotes the time evolution operator in perturbative form, incorporating the interaction with the fermionic spinor field $\Psi$. It is explicitly given by
\begin{eqnarray}
    \mathcal{\hat{U}}_{\Psi} = \mathcal{\hat{U}}^{(0)}_{\Psi} + \mathcal{\hat{U}}^{(1)}_{\Psi} + \mathcal{\hat{U}}^{(2)}_{\Psi} + \mathcal{O}(\lambda^3_{\Psi})~,
    \label{U}
\end{eqnarray}
being the perturbative terms, given by
\begin{align}
    \mathcal{\hat{U}}^{(0)}_{\Psi} &= \mathbb{I}~,\\[6pt]
    \mathcal{\hat{U}}^{(1)}_{\Psi} &= -i\lambda_{\Psi}\int_{-\infty}^{\infty}\d\tau\chi(\tau)\mu(\tau):\overline{\Psi}\Psi[x(\tau)]:~,\\[6pt]
    \mathcal{\hat{U}}^{(2)}_{\Psi} &= -\lambda^2_{\Psi}\int^{+\infty}_{-\infty} \d\tau \int^{+\tau}_{-\infty} \d\tau' \chi(\tau)\chi(\tau') \mu(\tau)\mu(\tau') :\overline{\Psi}\Psi[x(\tau)]: :\overline{\Psi}\Psi[x(\tau')]:~,
\end{align}
where $\mu(\tau) = [\hat{\sigma}_{+}\e^{i\Omega\tau} + \hat{\sigma}_{-}\e^{-i\Omega\tau}]$, this term can be physically interpreted as the detector’s response (or click) to the presence of the quantum field. Furthermore, the operators $\hat{\sigma}_{+} = |e\rangle\langle g|$ and $\hat{\sigma}_{-} = |g\rangle\langle e|$ are defined as the raising (excitation) and lowering (de-excitation) operators, respectively.

We now proceed to examine the final state of the UDW detector. To perform this analysis, it is necessary to trace over the degrees of freedom associated with the field. This procedure yields $\hat{\rho}^{\mathrm{out}}_{D,\Psi} = \mathrm{Tr}_{\vert0_{\mathcal{M}}\rangle}[\hat{\rho }^{\mathrm{out}}_{\Psi}]$.\\[2pt] 
After performing extensive and meticulous calculations, and defining the integrals as follows
\begin{align}
    \mathcal{B}^{\pm}_{\Psi} &= \int^{+\infty}_{-\infty} \d\tau \int^{+\infty}_{-\infty} \d\tau' \chi(\tau)\chi(\tau') \e^{\pm i\Omega(\tau+\tau')} \mathcal{W}_{\Psi}(\tau, \tau')~,\label{B+-}\\[8pt]
    \mathcal{G}^{\pm}_{\Psi} &= \int^{+\infty}_{-\infty} \d\tau \int^{+\tau}_{-\infty} \d\tau' \chi(\tau)\chi(\tau')  \e^{\pm i\Omega(\tau-\tau')} \mathcal{W}_{\Psi}(\tau, \tau')~,\label{intG}
    \end{align}
and, we omit the superscript ``$\mathrm{out}$'',  we obtain
\begin{equation}
    \hat{\rho}_{D,\Psi} = \renewcommand{\arraystretch}{1.25}\begin{pmatrix}
\hat{\rho}^{\mathrm{gg}}_{D,\Psi} & \hat{\rho}^{\mathrm{ge}}_{D,\Psi} \\
\hat{\rho}^{\mathrm{eg}}_{D,\Psi} & \hat{\rho}^{\mathrm{ee}}_{D,\Psi}
\end{pmatrix}
\label{final rho}
\end{equation}
where
\begin{eqnarray}
    \hat{\rho}^{\mathrm{gg}}_{D,\Psi} &=& \cos^2{\frac{\theta}{2}} + \lambda^2_{\Psi}\left[ \sin^2{\frac{\theta}{2}} \mathcal{F}^{+}_{\Psi} - 2\cos^2{\frac{\theta}{2}} \mathbf{Re}\left(\mathcal{G}^-_{\Psi}\right) \right]~, \label{rhogg}\\[6pt]
    \hat{\rho}^{\mathrm{ee}}_{D,\Psi} &=& \sin^2{\frac{\theta}{2}} + \lambda^2_{\Psi}\left[ \cos^2{\frac{\theta}{2}} \mathcal{F}^{-}_{\Psi} - 2\sin^2{\frac{\theta}{2}} \mathbf{Re}\left(\mathcal{G}^+_{\Psi}\right) \right]~, \label{rhoee}\\[6pt]
    \hat{\rho}^{\mathrm{eg}}_{D,\Psi} &=& \sin{\theta} \left[\frac{\e^{-i\varphi}}{2} + \frac{\lambda^2_{\Psi}}{2} \Big\{\e^{+i\varphi}\mathcal{B}^+_{\Psi} - \e^{-i\varphi} \left(\mathcal{G}^+_{\Psi} + \mathcal{G}^{-*}_{\Psi}\right) \Big\}\right]~,\label{rhoeg}\\[6pt]
    \hat{\rho}^{\mathrm{ge}}_{D,\Psi} &=& \sin{\theta} \left[\frac{\e^{+i\varphi}}{2} + \frac{\lambda^2_{\Psi}}{2} \Big\{\e^{-i\varphi}\mathcal{B}^-_{\Psi} - \e^{+i\varphi} \left(\mathcal{G}^-_{\Psi} + \mathcal{G}^{+*}_{\Psi}\right) \Big\}\right]~.\label{rhoge}
\end{eqnarray}
It is important to highlight that the terms proportional to $\lambda^{2}_{\Psi}$ cancel each other out; and therefore, by the unitary condition of the density matrix, we observe that
\begin{equation}
    \mathbf{Re}(\mathcal{G}^{-}_{\Psi}) = \frac{1}{2} \left( \mathcal{F}^{+}_{\Psi}\sin^2{\frac{\theta}{2}} + \mathcal{F}^{-}_{\Psi} \cos^2{\frac{\theta}{2}}\right)=\frac{\sigma}{2} \left( \overline{\mathcal{R}}^{+}_{\Psi}\sin^2{\frac{\theta}{2}} + \overline{\mathcal{R}}^{-}_{\Psi} \cos^2{\frac{\theta}{2}}\right)~,
    \label{e_60}
\end{equation}
where $\overline{\mathcal{R}}^{\pm} \equiv \mathcal{R}^{\pm}/\Omega$~.\\[4pt]
In a two-level quantum system, the degree of coherence between the two states can be quantified using the $l^1$ norm quantum coherence. This measure captures the effects arising from quantum superposition between the system's states. Specifically, for the case of our system described by the density matrix in eq.~(\ref{final rho}), the corresponding $l^1$ norm coherence is given by the sum of the absolute values of the off-diagonal elements \cite{Baumgratz2014QuantifyingCoherence}, leading to
\begin{equation}
    \mathcal{C}^{l^1}_{\Psi}(\hat{\rho}_{D,\,\Psi}) = \sum_{i \neq j} \mid \hat{\rho}_{D,\,\Psi}^{ij}\mid\,=\hat{\rho}_{D,\,\Psi}^{\mathrm{eg}}+\hat{\rho}_{D,\,\Psi}^{\mathrm{ge}}~.\label{e_61}
\end{equation}
\subsection{Massless case}
\noindent To evaluate the $l^{1}$ norm quantum coherence in the massless case, we first examine the off-diagonal elements of $\hat{\rho}_{D,\Psi,0}$ in the long interaction-time regime, i.e., for~$\sigma \gg 1$. Thus, taking the eqs. \eqref{rhoeg}, \eqref{rhoge}, and using eq.~\eqref{e_60}, we get
\begin{align}
\hat{\rho}^{\mathrm{eg}}_{D,\,\Psi,\,0} 
&= \frac{\sin\theta}{2} \left[ \e^{-i\varphi} - \sigma \lambda^{2}_{\Psi}\, \e^{-i\varphi} \left( \overline{\mathcal{R}}^{-}_{\Psi,\,0} \cos^{2}\frac{\theta}{2} + \overline{\mathcal{R}}^{+}_{\Psi,\,0} \sin^{2}\frac{\theta}{2} \right) \right]~,\label{e_62} \\[6pt]
\hat{\rho}^{\mathrm{ge}}_{D,\,\Psi,\,0} 
&= \frac{\sin\theta}{2} \left[ \e^{+i\varphi} - \sigma \lambda^{2}_{\Psi}\, \e^{+i\varphi} \left( \overline{\mathcal{R}}^{-}_{\Psi,\,0} \cos^{2}\frac{\theta}{2} + \overline{\mathcal{R}}^{+}_{\Psi,\,0} \sin^{2}\frac{\theta}{2} \right) \right]~.\label{e_63}
\end{align}
Substituting eqs.~\eqref{e_62}, \eqref{e_63}, into eq.~\eqref{e_61}, $l^{1}$-norm quantum coherence in the massless case takes the form
\begin{equation}
    \mathcal{C}_{\Psi,\,0}^{l^1} \approx  \vert\sin{\theta}\vert \Bigg[ 1 -\frac{\sigma\overline{\lambda}^2_{\Psi}}{60\pi^3} \frac{(1 + 5\overline{a}^2 + 4\overline{a}^4)}{(\e^{2\pi/\overline{a}}-1)}\Bigg( \cos^2{\frac{\theta}{2}} + \e^{2\pi/\overline{a}} \sin^2{\frac{\theta}{2}} \Bigg) \Bigg] + \mathcal{O}(\overline{\lambda}_{\Psi}^4)~.
    \label{e_64}
\end{equation}
Thus, we obtain $l^1$ norm quantum coherence for a qubit undergoing uniform acceleration that interacts with a massless fermionic spinor field for a long time. Note that for the coherence of an accelerated single-qubit linearly coupled with a scalar field $\mathcal{C}_{\phi,0}^{l^1}$, see refs.~\cite{barros2024detecting, pedro2025mitigating} taking the corresponding limits for this.
\subsection{Massive case}
\noindent For the massive coupling scenario, we calculate the $l^1$ norm quantum coherence for both small- and large-mass approximation in the limit $\sigma\gg1$.
\subsubsection{Small-mass approximation}
\noindent Under the small-mass approximation, the off-diagonal terms of $\hat{\rho}_{D,\,\Psi,\,m}$ becomes
\begin{align}
\hat{\rho}^{\mathrm{eg}}_{D,\,\Psi,\,sm} 
&= \frac{\sin\theta}{2} \left[ \e^{-i\varphi} - \sigma \lambda^{2}_{\Psi}\, \e^{-i\varphi} \left( \overline{\mathcal{R}}^{-}_{\Psi,\,sm} \cos^{2}\frac{\theta}{2} + \overline{\mathcal{R}}^{+}_{\Psi,\,sm} \sin^{2}\frac{\theta}{2} \right) \right],\label{e_65} \\[6pt]
\hat{\rho}^{\mathrm{ge}}_{D,\,\Psi,\,sm} 
&= \frac{\sin\theta}{2} \left[ \e^{+i\varphi} - \sigma \lambda^{2}_{\Psi}\, \e^{+i\varphi} \left( \overline{\mathcal{R}}^{-}_{\Psi,\,sm} \cos^{2}\frac{\theta}{2} + \overline{\mathcal{R}}^{+}_{\Psi,\,sm} \sin^{2}\frac{\theta}{2} \right) \right].\label{e_66}
\end{align}
Substituting eqs.~\eqref{e_65}, and  \eqref{e_66}, into eq.~\eqref{e_61}, $l^{1}$-norm of quantum coherence in the massless case takes the form
\begin{align}
    \mathcal{C}_{\Psi,\,sm}^{l^1} 
    &\approx \vert\sin{\theta}\vert \Bigg[ 1 -\frac{\sigma\overline{\lambda}^2_{\Psi}}{60\pi^3} \frac{1}{(\e^{2\pi/\overline{a}}-1)}\Big\{(1 + 5\overline{a}^2 + 4\overline{a}^4)-5\overline{m}^2(1 + \overline{a}^2) \Big\}  \nonumber\\
    &\times\Bigg( \cos^2{\frac{\theta}{2}} + \e^{2\pi/\overline{a}} \sin^2{\frac{\theta}{2}} \Bigg) \Bigg] + \mathcal{O}(\overline{\lambda}_{\Psi}^4)~.
    \label{e_67}
\end{align}

\subsubsection{Large-mass approximation}
\noindent Under the large-mass approximation, the off-diagonal terms of $\hat{\rho}_{D,\,\Psi,\,m}$ becomes
\begin{align}
\hat{\rho}^{\mathrm{eg}}_{D,\,\Psi,lm} 
&= \frac{\sin\theta}{2} \left[ \e^{-i\varphi} - \sigma \lambda^{2}_{\Psi}\, \e^{-i\varphi} \left( \overline{\mathcal{R}}^{-}_{\Psi,\,lm} \cos^{2}\frac{\theta}{2} + \overline{\mathcal{R}}^{+}_{\Psi,\,lm} \sin^{2}\frac{\theta}{2} \right) \right],\label{e_68} \\[6pt]
\hat{\rho}^{\mathrm{ge}}_{D,\,\Psi,lm} 
&= \frac{\sin\theta}{2} \left[ e^{+i\varphi} - \sigma \lambda^{2}_{\Psi}\, e^{+i\varphi} \left( \overline{\mathcal{R}}^{-}_{\Psi,\,lm} \cos^{2}\frac{\theta}{2} + \overline{\mathcal{R}}^{+}_{\Psi,\,lm} \sin^{2}\frac{\theta}{2} \right) \right].\label{e_69}
\end{align}
Substituting eqs.~\eqref{e_68}, \eqref{e_69}, into eq.~\eqref{e_61}, $l^{1}$ norm quantum coherence in the massless case takes the form
\begin{align}
    \mathcal{C}_{\Psi,\,lm}^{l^1} 
    &\approx \vert\sin{\theta}\vert \Bigg[ 1 -\frac{\sigma\,\overline{\lambda}^2_{\Psi}}{8\pi^2} \frac{\overline{m}^{2}}{(\e^{2\pi/\overline{a}}-1)}\Big\{1 + \overline{a}^2-12\,\overline{m}+48\,\overline{m}^{2}-64\,\overline{m}^{3}\Big\}   \nonumber\\
    &\times\Bigg( \cos^2{\frac{\theta}{2}} + \e^{2\pi/\overline{a}} \sin^2{\frac{\theta}{2}} \Bigg) \Bigg] + \mathcal{O}(\overline{\lambda}_{\Psi}^4)~.
    \label{e_70}
\end{align}
Thus, we obtain the $l^{1}$ norm quantum coherence for a uniformly accelerated qubit interacting with a massive fermionic spinor field, in both the small- and large-mass regimes, in the long interaction-time limit.

\section{\label{sec:level4}Numerical results}
In this section, we analyse our findings for both the massless and massive coupling scenarios.
\subsection{Massless case}
\begin{figure}[h!]
    \centering
    \includegraphics[width=0.6\linewidth]{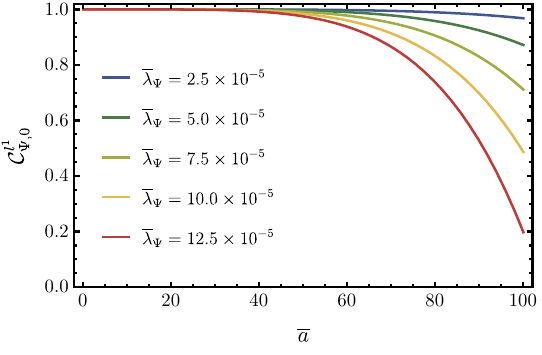}
    \caption{Quantum coherence $\mathcal{C}_{\Psi,0}^{l^1}$ as a function of the dimensionless parameter $\overline{a}$ for different values of $\overline{\lambda}_{\Psi}$. We fix the following parameters: $\theta = \pi/2$ and $\sigma = 10$.}
    \label{F2}
\end{figure}

\begin{figure}[h!]
    \centering
    \includegraphics[width=0.6\linewidth]{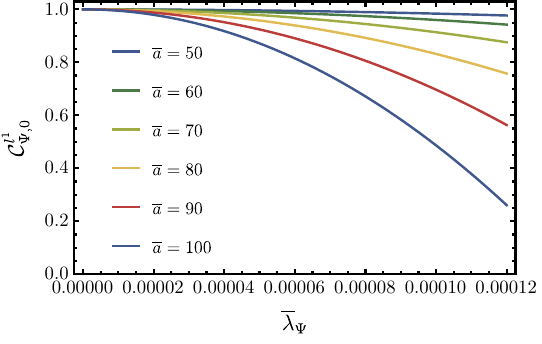}
    \caption{Quantum coherence $\mathcal{C}_{\Psi,0}^{l^1}$ as a function of the dimensionless parameter $\overline{\lambda}_{\Psi}$ for different values of $\overline{a}$. We fix the following parameters: $\theta = \pi/2$ and $\sigma = 10$.}
    \label{F3}
\end{figure}

As seen in \cite{takagi1985response, Takagi1986vacuum, Louko2016, hummer2016renormalized, gray2018scalar, mokhtar2020radiation, wu2023accelerating}, the accelerated detector coupled with a fermionic spinor field responds more strongly than a detector coupled linearly with the scalar field. In Figure~\ref{F2} and~\ref{F3}, we plot the quantum coherence $\mathcal{C}_{\Psi,\,0}^{l^1}$ as a function of the dimensionless parameters $\overline{a}$ and $\overline{\lambda}_{\Psi}$, respectively. These results show that as we increase the acceleration of qubit (Figure~\ref{F2}) or as we increase the coupling parameters (Figure~\ref{F3}), the quantum coherence is degraded more rapidly.

\begin{figure}[h!]
    \centering
    \includegraphics[width=0.6\linewidth]{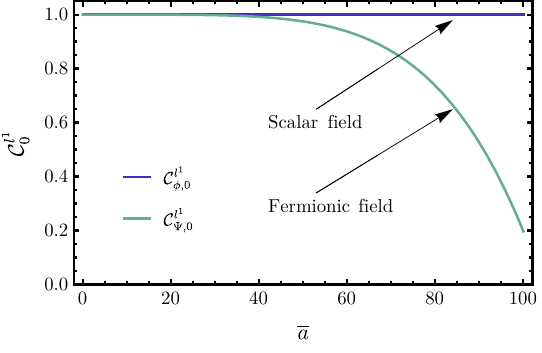}
    \caption{Comparison between quantum coherence for the case of the fermionic spinor field $\mathcal{C}_{\Psi,0}^{l^1}$ and for the case of the scalar field $\mathcal{C}_{\phi,0}^{l^1}$, both as a function of the dimensionless parameter $\overline{a}$. We fix the following parameters: $\theta = \pi/2$ and $\sigma = 10$.}
    \label{F4}
\end{figure}

In Figure~\ref{F4}, we plot a comparison between the quantum coherence of an accelerated detector coupled with the fermionic spinor field and with the scalar field, where by assuming that $\Omega = 1$ we can assume and choose that $\overline{\lambda}_{\Psi} = \lambda_{\phi} = 1.25 \times 10^{-3}$. Through this comparison it is possible to see that the degradation of coherence occurs much more quickly for coupling with the fermionic field when compared with coupling with the scalar field. Additionally, note that, due to the adopted coupling scale, the degradation of coherence is visually imperceptible.

\noindent Thus, a detector coupled to a fermonic field responds more due to the larger dimensionality of its coupling constant, which at $d = 4$ we have $\Delta_{\Psi} = 2$, contrasting with $\Delta _ {\phi} = 0$ for scalar fields~\cite{gray2018scalar}. This results in qualitatively distinct behavior and the presence of intrinsic ultraviolet differences in correlation calculations, indicating a response of greater magnitude or complexity~\cite{wu2023accelerating}. Note that it is important to emphasize the difference in coupling strength between the different types of detectors, namely: the coupling strength for the fermionic case is much greater when compared to the linear and quadratic cases of the scalar field \cite{barros2025quadratic}.

Additionally, a fermionic detector in $d$ dimensions behaves like a scalar detector in $2d$ dimensions (see the theorem in ref.~\cite{Louko2016}), suggesting an interaction with a richer phase space~\cite{gray2018scalar}. Finally, the fermionic detector response is linked to Helmholtz's free energy density, a more comprehensive thermodynamic property than the numerical density of bosons, to which scalar detectors respond \cite{mokhtar2020radiation}. Therefore, due to all this, according to the probability transition rate, consequently, coherence degradation also increases.

\subsection{Massive case}

\begin{figure}[h!]
    \centering
    \includegraphics[width=0.6\linewidth]{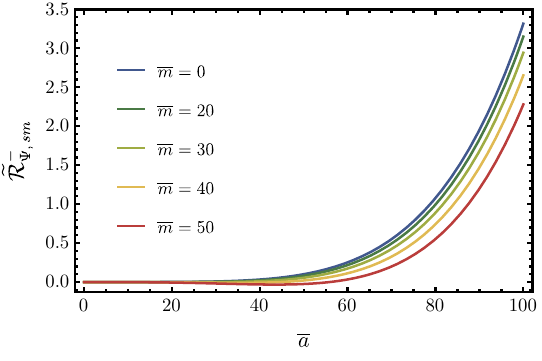}
    \caption{Excitation rate $\widetilde{\mathcal{R}}^{-}_{\Psi,\,sm}$ as a function of the parameter $\overline{a}$ for different values of the parameter $\overline{m}$. Where we fix the parameters: $\overline{\lambda}_{\Psi} = 1 \times 10^{-3}$ and $\sigma = 10$. Where $\widetilde{\mathcal{R}}^{\pm}_{\Psi,\,sm} \equiv \lambda_{\Psi}^{2}\overline{\mathcal{R}}^{\pm}_{\Psi,\,sm}$.}
    \label{F5}
\end{figure}

\begin{figure}[h!]
    \centering
    \includegraphics[width=0.6\linewidth]{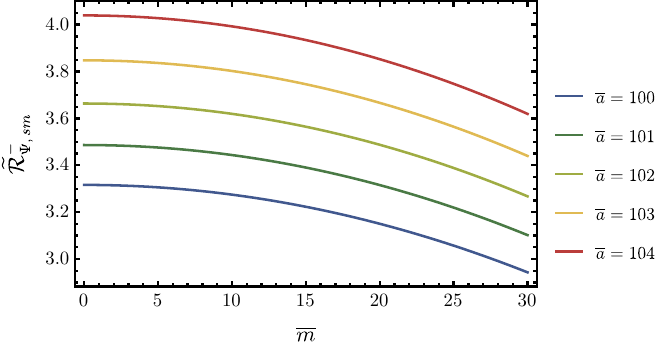}
    \caption{Excitation rate $\widetilde{\mathcal{R}}^{-}_{\Psi,\,sm}$ as a function of the parameter $\overline{m}$ for different values of the parameter $\overline{a}$. Where we fix the parameters: $\overline{\lambda}_{\Psi} = 1 \times 10^{-3}$ and $\sigma = 10$. Where $\overline{\mathcal{R}}^{-} \equiv \mathcal{R}^{-}/\Omega$. Where $\widetilde{\mathcal{R}}^{\pm}_{\Psi,\,sm} \equiv \lambda_{\Psi}^{2}\overline{\mathcal{R}}^{\pm}_{\Psi,\,sm}$.}
    \label{F6}
\end{figure}

In Figure~\ref{F5}, we analyze the excitation probability rate as a function of acceleration by varying the mass parameter. This plot shows the effects caused by the mass of the fermionic spinorial field. Note that as the mass parameter is increased, the excitation rate decreases.
Note that as the mass parameter is increased, the excitation rate decreases. Furthermore, in Figure~\ref{F6}, we plot the excitation rate as a function of the mass parameter for different acceleration values. Through the plots shown in Figure~\ref{F5} and Figure~\ref{F6}, it is possible to observe the effects caused by the mass of the fermionic spinorial field on the excitation probability rates of the accelerated detector. It is evident that the field mass is a parameter that can significantly reduce the detector response.

\noindent This finding is similar to that found in ref.~\cite{pedro2025mitigating} for the case of linear coupling with a massive scalar field. This means that the detector has difficulty absorbing massive particles, implying that the absorption of a single quantum of mass $m$ by the detector will become increasingly difficult as the mass approaches the spacing between energy levels $\Omega$. In other words, as the particle's ``rest energy'' approaches $\Omega$, the detector's response decreases. For more details see the refs.~\cite{birrell1984quantum, pedro2025mitigating}. Therefore, our findings show a similar effect, but for the fermionic field.

\begin{figure}[h!]
    \centering
    \includegraphics[width=0.6\linewidth]{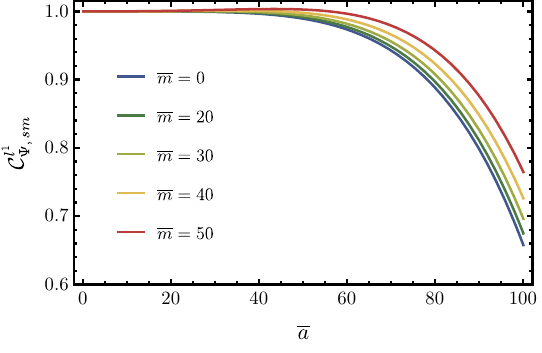}
    \caption{Coherence $\mathcal{C}^{l^1}_{\Psi,\, sm}$ as a function of $\overline{a}$ for different values of $\overline{m}$. Here we have kept the following parameters constant: $\overline{\lambda}_{\Psi} = 1 \times 10^{-4}$, $\sigma=10$ and $\theta=\pi/2$.}
    \label{F7}
\end{figure}

\begin{figure}[h!]
    \centering
    \includegraphics[width=0.6\linewidth]{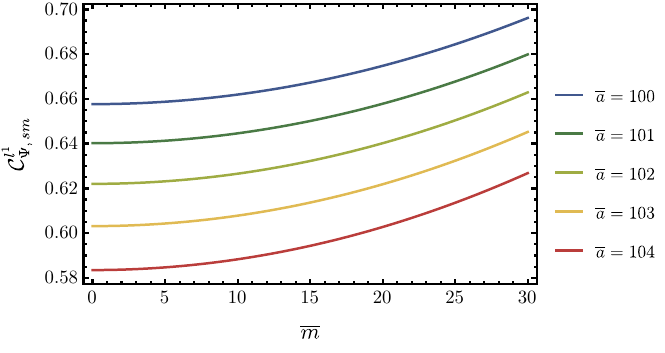}
    \caption{Coherence $\mathcal{C}^{l^1}_{\Psi,\, sm}$ as a function of $\overline{m}$ for different values of $\overline{a}$. Here we have kept the following parameters constant: $\overline{\lambda}_{\Psi} = 1 \times 10^{-4}$, $\sigma=10$ and $\theta=\pi/2$.}
    \label{F8}
\end{figure}

As shown in Figure~\ref{F7}, we plotted the quantum coherence of an accelerated single-qubit as a function of the acceleration parameter for different values of mass. Furthermore, in Figure~\ref{F8} we analyzed this same quantum coherence as a function of the mass parameter for different values of acceleration. Through these plots, it is easy to observe the effects of the mass parameter on quantum coherence. Notice that as we increase the mass, the degradation of quantum coherence is mitigated. This occurs because massive particles have difficulty being absorbed; that is, the detector responds less and less as the mass of the particles increases. As is well known, coherence degradation is caused by Unruh radiation (caused by acceleration) which is absorbed by the detector, causing changes in its internal states. Thus, if the detector absorbs fewer particles, then the coherence will also degrade less.

\noindent This effect also occurs in the case of linear coupling with a massive scalar field. However, for the fermionic field it is different because the fermionic coupling is stronger due to the fact that it has more interaction channels, which causes the acceleration radiation to produce more particles. Because of this, in the case of the fermionic field, the detector can absorb much larger particles when compared to the case of the scalar field, which only absorbs particles with mass $0 \leq \overline{m} < 1$~\cite{pedro2025mitigating}.

\section{\label{sec:level7}Conclusions}
In this work, we investigate the probability transition rates of a UDW detector and its applications in an accelerated single-qubit interacting (during a finite time) with a massless and massive fermionic spinorial field. First, we analytically obtained the transition rates for the fermionic field considering both the massless and massive coupling scenarios. Our findings shows the intriguing results for both the massless and massive coupling cases.

In case of massless ($\overline{m} = 0$) fermionic coupling scenario, we observe that the quantum coherence is degraded more rapidly when we increase the acceleration of the qubit and the coupling strength between the qubit and the surrounding fermionic quantum field. From our results, we can easily identify the effects caused solely by the coupling of the fermionic field when compared to the coupling with the scalar field. Through this comparison, it becomes evident that the degradation of coherence occurs significantly more rapidly for coupling to a fermionic field than for coupling to a scalar field. This occurs because the fermionic field exhibits more interaction channels than the scalar field, since the fermionic field behaves like a scalar field in 2d dimensions, suggesting greater interaction with phase space~\cite{gray2018scalar}. 

For the massive fermionic coupling scenario, we observe that the transition rate of the accelerated detector is reduced as we increase the mass. In other words, the detector has difficulty absorbing massive particles, meaning that the absorption of a single quantum of mass $m$ becomes progressively suppressed as the mass approaches the energy-level spacing $\Omega$. In other words, as the particle’s rest energy approaches $\Omega$, the response of the detector correspondingly diminishes. We also observe that when we increase the acceleration of the detector, excitation rate of the detector also increases, which is consistent with results of the usual scalar field coupling scenario.

Subsequently, when applying these transition rates to the accelerated qubit to find the well-known quantum coherence degradation due to the Unruh effect, we verified that mass causes the mitigation of coherence degradation. This effect also manifests in the case of linear coupling to a massive scalar field~\cite{pedro2025mitigating}. However, the situation differs for a fermionic field, as the fermionic coupling is stronger due to the presence of additional interaction channels, which leads the acceleration-induced radiation to produce a greater number of particles. Consequently, for a fermionic field, the detector is capable of absorbing substantially heavier particles compared to the scalar-field case, in which only particles with masses satisfying $0 \leq \overline{m} < 1$ are absorbed.

Therefore, our work shows that due to the fact that the scalar field and the fermionic field are quite different in many aspects, the information also behaves differently for the two couplings. Furthermore, we also verified that, in addition to the type of coupling, mass also plays a very important role in both cases; mass acts as a protective factor against Unruh radiation due to acceleration. These findings are valuable for the RQI community because they show ways to avoid information loss in accelerated quantum systems through the manipulation of the quantum field mass.
\begin{acknowledgments}
\noindent A.M would like to thank S. N. Bose National Centre for Basic Sciences for providing the financial support. P.H.M.B acknowledges the Brazilian funding agency CAPES for financial support through grant No.~88887.674765/2022-00 (Doctoral Fellowship – CAPES).
\end{acknowledgments}
\begin{appendices}
\section{Wightman function of the massive scalar field}\label{App:A}
\noindent In this appendix, we provide a brief derivation of the Wightman function for a massive scalar field, as given in eq.~\eqref{e_30}.

\noindent The Wightman function of the massive scalar field is given by
\begin{equation}
    \mathcal{W}_{\phi,\,m}(x,x')=\ev{\hat{\phi}(x)\,\hat{\phi}(x')}{0_M}~,~\label{e_A.1}
\end{equation}
with the usual mode expansion of the massive scalar field
\begin{equation}
    \phi(t,\vec{x})=\int_{-\infty}^{\infty} \frac{\d^3\vec{k}}{\sqrt{(2\pi)^{3}\,2\omega_{\vec{k}}}}\left(\hat{a}_{\vec{k}}\,\e^{-i\omega_{\vec{k}}t+i\vec{k}\cdot \vec{x}}+\text{h.c}\right)~\label{e_A.2}
\end{equation}
where $\omega_{\vec{k}}=\sqrt{k^2+m^2}$.\\[2pt] 
Substituting eq.~\eqref{e_A.2} into eq.~\eqref{e_A.1}, doing some mathematical manipulation, we obtain
\begin{equation}
    \mathcal{W}_{\phi,\,m}(x,x')=\lim_{\epsilon\rightarrow0^+}\frac{1}{(2\pi)^3}\int_{-\infty}^{\infty}\frac{\d^3\vec{k}}{2\omega_{\vec{k}}}\,\e^{-i\omega_{\vec{k}}T+i\vec{k}\cdot \vec{X}}
\end{equation}
with $T\equiv t-t'-i\epsilon\,,\,\,\vec{X}\equiv\vec{x}-\vec{x}'$, and $X=|\vec{X}|$.\\[2pt] 
Changing the integral measure into spherical polar coordinates, and after computing the angular part, we obtain,
\begin{align}
    \mathcal{W}_{\phi,\,m}(x,x')
    &=\frac{1}{4\pi^2X}\int_{0}^{\infty}\d k\frac{k}{\sqrt{k^2+m^2}}\sin{(kX)}\,\e^{-i\sqrt{k^2+m^2}\,T}\nonumber\\
    &\equiv\frac{1}{4\pi^2X}\,I(m,X,T)~.
\end{align}
Substituting $k=m\sinh{u}$, we obtain
\begin{align}
    I(m,X,T)&=m\int_{0}^{\infty}\d u\,\sinh{u}\,\sin{(mX\sinh{u})}\,\e^{-imT\,\cosh{u}}\nonumber\\
    &=-\frac{\partial}{\partial X}\left[\int_{0}^{\infty}\d u\,\e^{-imT\,\cosh{u}}\,\cos{(mX\sinh{u})}\, \right]\nonumber\\
    &=-\frac{\partial}{\partial X} \mathcal{K}_{0}(m\,s)\quad \text{where\,\,} s\equiv\sqrt{-(T^2-X^2)}~.
\end{align}
Now using the properties of the modified Bessel function of the second kind of zeroth order $\mathcal{K}_{0}(z)$ \cite{abramowitz1964handbook},
\begin{equation}
    \frac{\d}{\d z}\mathcal{K}_{0}(z)=-\mathcal{K}_{1}(z)
\end{equation}
we get
\begin{equation}
    I(m,X,T)=\frac{mX}{s}\mathcal{K}_{1}(ms)~.
\end{equation}
Hence, Wightman function of the massive scalar field becomes,
\begin{equation}
    \mathcal{W}_{\phi,\,m}(x,x')=\frac{m}{4\pi^2}\left[\frac{\mathcal{K}_{1}(ms)}{s}\right]=\frac{m}{4\pi^2}\left[\frac{\mathcal{K}_{1}(m\sqrt{-\rho})}{\sqrt{-\rho}}\right]~.
\end{equation}
This form is given in eq.~\eqref{e_30}.
\section{Wightman function of the massive fermionic field}\label{App:B}
\noindent In this appendix, we provide a brief derivation of the Wightman function for a massive fermionic field, as given in eq.~\eqref{e_31}.\\[2pt]
From eq.~\eqref{e_29}, we have seen that the Wightman function for a massive fermionic field can be obtained directly from the massive scalar Wightman function given in eq.~\eqref{e_30}.\\[2pt]
Let's consider the massive scalar Wightman function in the form
\begin{equation}
    \mathcal{W}_{\phi,\,m}(x,x')=\frac{m}{4\pi^2}\left[\frac{\mathcal{K}_{1}(ms)}{s}\right]\equiv\frac{1}{4\pi^2}f(s)~\label{e_B.1}
\end{equation}
where $s=\sqrt{-\rho}$ and $\rho=(t-t'-i\epsilon)^2-|\vec{x}-\vec{x}'|^2$.\\
In a four-vector notation this $\rho$ can be written as
\begin{equation}
    \rho=\eta_{\alpha\beta}\,(x^{\alpha}-x'^{\,\alpha})(x^{\beta}-x'^{\,\beta})-i\epsilon+\mathcal{O}(\epsilon^2)~\label{e_B.2}
\end{equation}
where $\alpha,\beta$ both takes the value $0,1,2,3$.\\[2pt]
Now computing the following derivative $\partial_{\mu}\mathcal{W}_{\phi,\,m}(x,x')$ we get
\begin{equation}
    \partial_{x^{\mu}}\mathcal{W}_{\phi,\,m}(x,x')=\frac{1}{4\pi^2}\frac{\partial f(s)}{\partial s}\frac{\partial s}{\partial x^{\mu}}~.\label{e_B.3}
\end{equation}
After doing some straightforward calculation, we obtain the following results
\begin{align}
    \frac{\partial f(s)}{\partial s}&=-\frac{m^2}{s}\mathcal{K}_{2}(ms)~\label{e_B.4}\\
    \frac{\partial s}{\partial x^{\mu}}&=-\frac{(x_{\mu}-x'_{\,\mu})}{s}~.\label{e_B.5}
    \end{align}
Hence, using the above results, we get
\begin{equation}
    \partial_{x^{\mu}}\mathcal{W}_{\phi,\,m}(x,x')=\frac{1}{4\pi^2}\frac{m^2}{s^2}\mathcal{K}_{2}(ms)(x_{\mu}-x'_{\,\mu})~.\label{e_B.6}
\end{equation}
Similarly, we get
\begin{equation}
    \partial_{x'^{\nu}}\mathcal{W}_{\phi,\,m}(x,x')=-\frac{1}{4\pi^2}\frac{m^2}{s^2}\mathcal{K}_{2}(ms)(x_{\nu}-x'_{\,\nu})~.\label{e_B.7}
\end{equation}
Inserting eqs.~\eqref{e_B.1}, \eqref{e_B.6}, \eqref{e_B.7} into eq.~\eqref{e_29}, and doing some mathematical simplification; finally, the Wightman function for massive fermionic field turns out to be
\begin{equation}
    \mathcal{W}_{\Psi,\,m}(x,x')=\frac{m}{4\pi^{4}\rho}\bigg[\mathcal{K}^{2}_{1}\left(m\sqrt{-\rho}\right)-\mathcal{K}^{2}_{2}\left(m\sqrt{-\rho}\right)\bigg]~.\label{e_B.8}
\end{equation}
This form we have used in eq.~\eqref{e_31}.
\end{appendices}


\bibliographystyle{JHEP.bst}
\bibliography{main.bib}

@article{unruh1976,
  title = {Notes on black-hole evaporation},
  author = {Unruh, W. G.},
  journal = {Phys. Rev. D},
  volume = {14},
  issue = {4},
  pages = {870--892},
  numpages = {0},
  year = {1976},
  month = {Aug},
  publisher = {American Physical Society},
  doi = {10.1103/PhysRevD.14.870},
  url = {https://link.aps.org/doi/10.1103/PhysRevD.14.870}
}

@article{fulling1973,
  title = {Nonuniqueness of Canonical Field Quantization in Riemannian Space-Time},
  author = {Fulling, S. A.},
  journal = {Phys. Rev. D},
  volume = {7},
  issue = {10},
  pages = {2850--2862},
  numpages = {0},
  year = {1973},
  month = {May},
  publisher = {American Physical Society},
  doi = {10.1103/PhysRevD.7.2850},
  url = {https://link.aps.org/doi/10.1103/PhysRevD.7.2850}
}

@article{Davies_1975,
doi = {10.1088/0305-4470/8/4/022},
url = {https://dx.doi.org/10.1088/0305-4470/8/4/022},
year = {1975},
month = {apr},
publisher = {},
volume = {8},
number = {4},
pages = {609},
author = {Davies, P. C. W.},
title = {Scalar production in Schwarzschild and Rindler metrics},
journal = {J. Phys. A Math. Gen.}
}

@article{grove1983notes,
  doi = {10.1088/0305-4470/16/16/029},
url = {https://doi.org/10.1088/0305-4470/16/16/029},
year = {1983},
month = {nov},
publisher = {},
volume = {16},
number = {16},
pages = {3905},
author = {P G Grove and A C Ottewill},
title = {Notes on 'particle detectors'},
journal = {Journal of Physics A: Mathematical and General}
}

@book{wald1994quantum,
  author    = {Wald, Robert M.},
  title     = {Quantum Field Theory in Curved Spacetime and Black Hole Thermodynamics},
  publisher = {University of Chicago Press},
  year      = {1994},
  address   = {Chicago},
  isbn      = {978-0226870274},
  series    = {Chicago Lectures in Physics} 
}

@book{lancaster2014quantum,
  author = {Lancaster, Tom and Blundell, Stephen J.},
    title = {Quantum Field Theory for the Gifted Amateur},
    publisher = {Oxford University Press},
    year = {2014},
    month = {04},
    isbn = {9780199699322},
    doi = {10.1093/acprof:oso/9780199699322.001.0001},
    url = {https://doi.org/10.1093/acprof:oso/9780199699322.001.0001},
}

@article{Crispino2008,
  title = {The Unruh effect and its applications},
  author = {Crispino, L. C. B. and Higuchi, A. and Matsas, G. E. A.},
  journal = {Rev. Mod. Phys.},
  volume = {80},
  issue = {3},
  pages = {787--838},
  numpages = {0},
  year = {2008},
  month = {Jul},
  publisher = {American Physical Society},
  doi = {10.1103/RevModPhys.80.787},
  url = {https://link.aps.org/doi/10.1103/RevModPhys.80.787}
}

@inbook{DeWitt1980,
author = {DeWitt, B. S.},
title = {Quantum Gravity: The new synthesis},
booktitle = {General Relativity: An Einstein Centenary Survey},
pages = {680--745},
url = {https://ui.adsabs.harvard.edu/abs/2010grae.book.....H},
publisher = {Cambridge University Press},
year = {1980}
}

@inbook{davies1984,
author    = {Davies, P. C. W.},
  title     = {Particles do not exist},
  booktitle = {{Quantum theory of gravity}: {Essays in honor of the 60th birthday of Bryce S. DeWitt}},
  editor    = {{Christensen, Steven M.}},
  publisher = {Adam Hilger Ltd.},
  year      = {1984},
  pages     = {66--77},
  address   = {Bristol, UK}
}

@article{martin2011relativistic,
  title={Relativistic Quantum Information: developments in Quantum Information in general relativistic scenarios},
  author={Mart{\'\i}n-Mart{\'\i}nez, E.},
  journal={arXiv preprint arXiv:1106.0280},
  year={2011}
}

@article{Mann2012,
doi = {10.1088/0264-9381/29/22/220301},
url = {https://dx.doi.org/10.1088/0264-9381/29/22/220301},
year = {2012},
month = {nov},
publisher = {},
volume = {29},
number = {22},
pages = {220301},
author = {R. B. Mann and T. C. Ralph},
title = {Relativistic quantum information},
journal = {Class. Quantum Grav.}
}

@incollection{hawking1975particle,
  title={Particle creation by black holes},
  author={Hawking, Stephen W},
  booktitle={Euclidean quantum gravity},
  pages={167--188},
  year={1975},
  publisher={World Scientific},
  doi={10.1007/BF02345020},
  url={https://doi.org/10.1007/BF02345020}
}

@article{Adlam2015,
  title = {Device-independent relativistic quantum bit commitment},
  author = {Adlam, E. and Kent, A.},
  journal = {Phys. Rev. A},
  volume = {92},
  issue = {2},
  pages = {022315},
  numpages = {9},
  year = {2015},
  month = {Aug},
  publisher = {American Physical Society},
  doi = {10.1103/PhysRevA.92.022315},
  url = {https://link.aps.org/doi/10.1103/PhysRevA.92.022315}
}

@article{Landulfo2016,
  title = {Nonperturbative approach to relativistic quantum communication channels},
  author = {Landulfo, A. G. S.},
  journal = {Phys. Rev. D},
  volume = {93},
  issue = {10},
  pages = {104019},
  numpages = {13},
  year = {2016},
  month = {May},
  publisher = {American Physical Society},
  doi = {10.1103/PhysRevD.93.104019},
  url = {https://link.aps.org/doi/10.1103/PhysRevD.93.104019}
}

@article{Martinez2015,
  title = {Harvesting correlations from the quantum vacuum},
  author = {Pozas-Kerstjens, A. and Mart\'{\i}n-Mart\'{\i}nez, E.},
  journal = {Phys. Rev. D},
  volume = {92},
  issue = {6},
  pages = {064042},
  numpages = {18},
  year = {2015},
  month = {Sep},
  publisher = {American Physical Society},
  doi = {10.1103/PhysRevD.92.064042},
  url = {https://link.aps.org/doi/10.1103/PhysRevD.92.064042}
}

@article{Martinez2016,
  title = {Entanglement harvesting from the electromagnetic vacuum with hydrogenlike atoms},
  author = {Pozas-Kerstjens, A. and Mart\'{\i}n-Mart\'{\i}nez, E.},
  journal = {Phys. Rev. D},
  volume = {94},
  issue = {6},
  pages = {064074},
  numpages = {27},
  year = {2016},
  month = {Sep},
  publisher = {American Physical Society},
  doi = {10.1103/PhysRevD.94.064074},
  url = {https://link.aps.org/doi/10.1103/PhysRevD.94.064074}
}

@article{Martinez2020,
  title = {Transmission of quantum information through quantum fields},
  author = {Simidzija, P. and Ahmadzadegan, A. and Kempf, A. and Mart\'{\i}n-Mart\'{\i}nez, E.},
  journal = {Phys. Rev. D},
  volume = {101},
  issue = {3},
  pages = {036014},
  numpages = {21},
  year = {2020},
  month = {Feb},
  publisher = {American Physical Society},
  doi = {10.1103/PhysRevD.101.036014},
  url = {https://link.aps.org/doi/10.1103/PhysRevD.101.036014}
}

@article{Martinez2021,
  title = {When entanglement harvesting is not really harvesting},
  author = {Tjoa, E. and Mart\'{\i}n-Mart\'{\i}nez, E.},
  journal = {Phys. Rev. D},
  volume = {104},
  issue = {12},
  pages = {125005},
  numpages = {21},
  year = {2021},
  month = {Dec},
  publisher = {American Physical Society},
  doi = {10.1103/PhysRevD.104.125005},
  url = {https://link.aps.org/doi/10.1103/PhysRevD.104.125005}
}

@article{Tjoa2022,
  title = {Quantum teleportation with relativistic communication from first principles},
  author = {Tjoa, E.},
  journal = {Phys. Rev. A},
  volume = {106},
  issue = {3},
  pages = {032432},
  numpages = {14},
  year = {2022},
  month = {Sep},
  publisher = {American Physical Society},
  doi = {10.1103/PhysRevA.106.032432},
  url = {https://link.aps.org/doi/10.1103/PhysRevA.106.032432}
}

@article{Lapponi2023,
  title = {Relativistic quantum communication between harmonic oscillator detectors},
  author = {Lapponi, A. and Moustos, D. and Bruschi, D. E. and Mancini, S.},
  journal = {Phys. Rev. D},
  volume = {107},
  issue = {12},
  pages = {125010},
  numpages = {19},
  year = {2023},
  month = {Jun},
  publisher = {American Physical Society},
  doi = {10.1103/PhysRevD.107.125010},
  url = {https://link.aps.org/doi/10.1103/PhysRevD.107.125010}
}

@article{Unruh1981Experimental,
  title = {Experimental Black-Hole Evaporation?},
  author = {Unruh, W. G.},
  journal = {Phys. Rev. Lett.},
  volume = {46},
  issue = {21},
  pages = {1351--1353},
  numpages = {0},
  year = {1981},
  month = {May},
  publisher = {American Physical Society},
  doi = {10.1103/PhysRevLett.46.1351},
  url = {https://link.aps.org/doi/10.1103/PhysRevLett.46.1351}
}

@article{tian2012unruh,
  author = {Tian, Z. and Jing, J.},
    title = {How the Unruh effect affects transition between classical and quantum decoherences},
    eprint = {1203.6141},
    archivePrefix = {arXiv},
    primaryClass = {quant-ph},
    doi = {10.1016/j.physletb.2011.12.064},
    pages = {264},
    volume = {707},
    journal = {Phys. Lett. B},
    year = {2012}
}

@article{Wang2016,
  title = {Irreversible degradation of quantum coherence under relativistic motion},
  author = {Wang, J. and Tian, Z. and Jing, J. and Fan, H.},
  journal = {Phys. Rev. A},
  volume = {93},
  pages = {062105},
  numpages = {6},
  year = {2016},
  month = {Jun},
  publisher = {American Physical Society},
  doi = {10.1103/PhysRevA.93.062105},
  url = {https://link.aps.org/doi/10.1103/PhysRevA.93.062105}
}

@article{he2018multipartite,
  author = {He, Juan and Ding, Zhi-Yong and Shi, Jia-Dong and Wu, Tao},
title = {Multipartite Quantum Coherence and Distribution under the Unruh Effect},
journal = {Annalen der Physik},
volume = {530},
number = {9},
pages = {1800167},
doi = {https://doi.org/10.1002/andp.201800167},
url = {https://onlinelibrary.wiley.com/doi/abs/10.1002/andp.201800167},
eprint = {https://onlinelibrary.wiley.com/doi/pdf/10.1002/andp.201800167},
year = {2018}
}

@article{Nesterov2020,
  title = {Decoherence as a detector of the Unruh effect},
  author = {Nesterov, A. I. and Fern\'andez, M. A. R. and Berman, G. P. and Wang, X.},
  journal = {Phys. Rev. Res.},
  volume = {2},
  issue = {4},
  pages = {043230},
  numpages = {5},
  year = {2020},
  month = {Nov},
  publisher = {American Physical Society},
  doi = {10.1103/PhysRevResearch.2.043230},
  url = {https://link.aps.org/doi/10.1103/PhysRevResearch.2.043230}
}

@Article{Zhang2022,
author = {Zhang, W. and Liu, X.},
title = {Quantum coherence of a circularly accelerated atom in a spacetime with a reflecting boundary},
journal = {Sci. Rep.},
year = {2022},
volume = {12},
number = {12577},
pages = {2045-2322},
doi = {10.1038/s41598-022-16647-9}
}

@article{huang2022,
  author={Huang, Zhiming},
title={Coherence behaviors of an atom immersing in a massive scalar field},
journal={The European Physical Journal D},
year={2022},
month={Apr},
day={16},
volume={76},
number={4},
pages={67},
issn={1434-6079},
doi={10.1140/epjd/s10053-022-00399-2},
url={https://doi.org/10.1140/epjd/s10053-022-00399-2}
}

@article{Harikrishnan2022,
  title = {Accessible and inaccessible quantum coherence in relativistic quantum systems},
  author = {Harikrishnan, S. and Jambulingam, S. and Rohde, P. P. and Radhakrishnan, C.},
  journal = {Phys. Rev. A},
  volume = {105},
  issue = {5},
  pages = {052403},
  numpages = {17},
  year = {2022},
  month = {May},
  publisher = {American Physical Society},
  doi = {10.1103/PhysRevA.105.052403},
  url = {https://link.aps.org/doi/10.1103/PhysRevA.105.052403}
}

@article{xu2023decoherence,
  author={Xu, Hao},
title={Decoherence and thermalization of Unruh-DeWitt detector in arbitrary dimensions},
journal={Journal of High Energy Physics},
year={2023},
month={Mar},
day={24},
volume={2023},
number={3},
pages={179},
issn={1029-8479},
doi={10.1007/JHEP03(2023)179},
url={https://doi.org/10.1007/JHEP03(2023)179}
}

@article{barros2024dispersive,
  title={Dispersive vacuum as a decoherence amplifier of an Unruh--DeWitt detector},
  author={Barros, P. H. M. and Costa, H. A. S.},
  journal={J. Phys. A: Math. Theor.},
  volume={57},
  url={https://iopscience.iop.org/article/10.1088/1751-8121/ad860b},
  number={44},
  pages={445305},
  year={2024},
  publisher={IOP Publishing}
}

@article{Takagi1986vacuum,
    author = {Takagi, S.},
    title = {Vacuum Noise and Stress Induced by Uniform Acceleration: Hawking-Unruh Effect in Rindler Manifold of Arbitrary Dimension},
    journal = {Prog. Theor. Phys. Suppl.},
    volume = {88},
    pages = {1-142},
    year = {1986},
    month = {03},
    issn = {0375-9687},
    doi = {10.1143/PTP.88.1},
    url = {https://doi.org/10.1143/PTP.88.1}
}

@article{Louko2016,
  title = {Unruh-DeWitt detector's response to fermions in flat spacetimes},
  author = {Louko, J. and Toussaint, V.},
  journal = {Phys. Rev. D},
  volume = {94},
  issue = {6},
  pages = {064027},
  numpages = {17},
  year = {2016},
  month = {Sep},
  publisher = {American Physical Society},
  doi = {10.1103/PhysRevD.94.064027},
  url = {https://link.aps.org/doi/10.1103/PhysRevD.94.064027}
}

@article{gray2018scalar,
  title={Scalar and fermionic Unruh Otto engines},
  author={Gray, F. and Mann, R. B.},
  journal={J. High Energ. Phys.},
  volume={2018},
  number={11},
  pages={1--34},
  year={2018},
  url={https://doi.org/10.1007/JHEP11(2018)174},
  publisher={Springer}
}

@article{Wightman1956,
  title = {Quantum Field Theory in Terms of Vacuum Expectation Values},
  author = {Wightman, A. S.},
  journal = {Phys. Rev.},
  volume = {101},
  issue = {2},
  pages = {860--866},
  numpages = {0},
  year = {1956},
  month = {Jan},
  publisher = {American Physical Society},
  doi = {10.1103/PhysRev.101.860},
  url = {https://link.aps.org/doi/10.1103/PhysRev.101.860}
}

@article{letaw1981quantized,
  title={Quantized scalar field in the stationary coordinate systems of flat spacetime},
  author={Letaw, J. R. and Pfautsch, J. D.},
  journal={Phys. Rev. D},
  volume={24},
  number={6},
  pages={1491},
  year={1981},
  url = {https://link.aps.org/doi/10.1103/PhysRevD.24.1491},
  publisher={APS}
}

@article{padmanabhan1982general,
  title={General covariance, accelarated frames and the particle concept},
  author={Padmanabhan, T.},
  journal={Astrophys. Space. Sci.},
  volume={83},
  pages={247--268},
  year={1982},
  url={https://doi.org/10.1007/BF00648558},
  publisher={Springer}
}

@article{rindler1966kruskal,
  title={Kruskal space and the uniformly accelerated frame},
  author={Rindler, W.},
  journal={Am. J. Phys},
  volume={34},
  number={12},
  url={https://doi.org/10.1119/1.1972547},
  pages={1174--1178},
  year={1966}
}

@article{kubo1957statistical,
  title={Statistical-mechanical theory of irreversible processes. I. General theory and simple applications to magnetic and conduction problems},
  author={Kubo, R.},
  journal={J. Phys. Soc. Jap.},
  volume={12},
  number={6},
  pages={570--586},
  year={1957},
  url={https://doi.org/10.1143/JPSJ.12.570},
  publisher={The Physical Society of Japan}
}

@article{Martin1959Schwinger,
  title = {Theory of Many-Particle Systems. I},
  author = {Martin, P. C. and Schwinger, J.},
  journal = {Phys. Rev.},
  volume = {115},
  issue = {6},
  pages = {1342--1373},
  numpages = {0},
  year = {1959},
  month = {Sep},
  publisher = {American Physical Society},
  doi = {10.1103/PhysRev.115.1342},
  url = {https://link.aps.org/doi/10.1103/PhysRev.115.1342}
}

@article{sriramkumar1996finite,
  doi = {10.1088/0264-9381/13/8/005},
url = {https://doi.org/10.1088/0264-9381/13/8/005},
year = {1996},
month = {aug},
publisher = {},
volume = {13},
number = {8},
pages = {2061},
author = {L Sriramkumar and T Padmanabhan},
title = {Finite-time response of inertial and uniformly accelerated Unruh-DeWitt detectors},
journal = {Classical and Quantum Gravity}
}

@article{Svaiter1992Inertial,
  title = {Inertial and noninertial particle detectors and vacuum fluctuations},
  author = {Svaiter, B. F. and Svaiter, N. F.},
  journal = {Phys. Rev. D},
  volume = {46},
  issue = {12},
  pages = {5267--5277},
  numpages = {0},
  year = {1992},
  month = {Dec},
  publisher = {American Physical Society},
  doi = {10.1103/PhysRevD.46.5267},
  url = {https://link.aps.org/doi/10.1103/PhysRevD.46.5267}
}

@article{Higuchi1993Uniformly,
  title = {Uniformly accelerated finite-time detectors},
  author = {Higuchi, A. and Matsas, G. E. A. and Peres, C. B.},
  journal = {Phys. Rev. D},
  volume = {48},
  issue = {8},
  pages = {3731--3734},
  numpages = {0},
  year = {1993},
  month = {Oct},
  publisher = {American Physical Society},
  doi = {10.1103/PhysRevD.48.3731},
  url = {https://link.aps.org/doi/10.1103/PhysRevD.48.3731}
}

@Article{Pedro2024robustness,
author = {Barros, P. H. M. and da Paz, I. G. and Neto, O. P. de Sá and Costa, H. A. S.},
title = {Robustness of Wave–Particle Duality under Unruh Effect},
journal = {Entropy},
volume = {26},
year = {2024},
number = {1},
url = {https://www.mdpi.com/1099-4300/26/1/1},
PubMedID = {38275481},
ISSN = {1099-4300},
doi = {10.3390/e26010001}
}

@inproceedings{jazaeri2019review,
  title={A review on quantum computing: From qubits to front-end electronics and cryogenic MOSFET physics},
  author={Jazaeri, F. and Beckers, A. and Tajalli, A. and Sallese, J. -M.},
  booktitle={2019 MIXDES-26th International Conference" Mixed Design of Integrated Circuits and Systems"},
  pages={15--25},
  year={2019},
  url={https://doi.org/10.23919/MIXDES.2019.8787164},
  organization={IEEE}
}

@incollection{kasirajan2021quantum,
  author = {Kasirajan, Venkateswaran},
  title = {The Quantum Superposition Principle and Bloch Sphere Representation},
  editor = {Kasirajan, Venkateswaran},
  booktitle = {Fundamentals of Quantum Computing: Theory and Practice},
  year = {2021},
  publisher = {Springer},
  address = {Cham},
  pages = {75--104},
  doi = {10.1007/978-3-030-63689-0_3},
  url = {doi.org}
}

@article{Baumgratz2014QuantifyingCoherence,
  title = {Quantifying Coherence},
  author = {Baumgratz, T. and Cramer, M. and Plenio, M. B.},
  journal = {Phys. Rev. Lett.},
  volume = {113},
  issue = {14},
  pages = {140401},
  numpages = {5},
  year = {2014},
  month = {Sep},
  publisher = {American Physical Society},
  doi = {10.1103/PhysRevLett.113.140401},
  url = {https://link.aps.org/doi/10.1103/PhysRevLett.113.140401}
}

@article{barros2024detecting,
  title={Detecting gravitational waves via coherence degradation induced by the Unruh effect},
  author={Barros, P. H. M. and Costa, H. A. S.},
  journal={Eur. Phys. J. C},
  volume={84},
  number={12},
  pages={1--11},
  year={2024},
  url={https://link.springer.com/article/10.1140/epjc/s10052-024-13639-z},
  publisher={Springer}
}

@article{pedro2025mitigating,
  title = {Mitigating the information degradation in a massive Unruh-DeWitt theory},
  author = {Barros, P. H. M. and Lima, F. C. E. and Almeida, C. A. S. and Costa, H. A. S.},
  journal = {J. High Energ. Phys.},
  volume = {2025},
  issue = {4},
  pages = {165},
  number = {},
  year = {2025},
  month = {Apr},
  publisher = {Spinger},
  doi = {10.1007/JHEP04(2025)165},
  issn = {1029-8479},
  url = {https://doi.org/10.1007/JHEP04(2025)165}
}

@article{takagi1985response,
  title={On the response of a rindler particle detector. iii},
  author={Takagi, S.},
  journal={Prog. Theor. Phys.},
  volume={74},
  number={3},
  url={https://doi.org/10.1143/PTP.74.501},
  pages={501--510},
  year={1985},
  publisher={Oxford University Press}
}

@article{mokhtar2020radiation,
  title={Radiation from a receding mirror: Unruh-DeWitt detector distinguishes a Dirac fermion from a scalar boson},
  author={Mokhtar, W. M. H. W.},
  journal={Class. Quantum Grav.},
  volume={37},
  url={https://doi.org/10.1088/1361-6382/ab6f0e},
  number={7},
  pages={075011},
  year={2020},
  publisher={IOP Publishing}
}

@article{wu2023accelerating,
  title={Accelerating Unruh-DeWitt detectors coupled with a spinor field},
  author={Wu, Dawei and Tang, Shan-Chang and Shi, Yu},
  journal={J. High Energ. Phys.},
  volume={2023},
  number={6},
  url={https://doi.org/10.1007/JHEP06(2023)190},
  pages={1--28},
  year={2023},
  publisher={Springer}
}

@article{barros2025quadratic,
  title={On the information behavior from quadratically coupled accelerated detectors},
  author={Barros, P. H. M. and Carvalho, P. R. S. and Costa, H. A. S.},
  journal={arXiv preprint arXiv:2505.14915},
  doi={https://doi.org/10.48550/arXiv.2505.14915},
  year={2025}
}

@book{birrell1984quantum,
  title={Quantum fields in curved space},
  author={Birrell, Nicholas David and Davies, PCW},
  year={1984},
  publisher={Cambridge university press},
  doi={10.1017/CBO9780511622632},
  url={https://www.cambridge.org/core/books/quantum-fields-in-curved-space/95376B0CAD78EE767FCD6205F8327F4C}
}

@article{alkofer2016quantum,
  title = {Quantum gravity signatures in the Unruh effect},
  author = {Alkofer, Natalia and D'Odorico, Giulio and Saueressig, Frank and Versteegen, Fleur},
  journal = {Phys. Rev. D},
  volume = {94},
  issue = {10},
  pages = {104055},
  numpages = {20},
  year = {2016},
  month = {Nov},
  publisher = {American Physical Society},
  doi = {10.1103/PhysRevD.94.104055},
  url = {https://link.aps.org/doi/10.1103/PhysRevD.94.104055}
}

@book{abramowitz1964handbook,
author = {Abramowitz, M. and Stegun, I.A.},
title= {Handbook of mathematical functions with formulas, graphs, and mathematical tables},
year = {1964},
address = {New York},
publisher = {Dover},
}

@book{gradshteyn2014table,
      author= "Gradshteyn, Izrail Solomonovich and Ryzhik, I M and Zwillinger, Daniel and Moll, Victor",
      title= "{Table of integrals, series, and products; 8th ed.}",
      publisher= "Academic Press",
      address= "Amsterdam",
      month= "Sep",
      year= "2014",
      url= "https://cds.cern.ch/record/1702455",
      doi= "0123849330",
}

@article{liu2025does,
  title={Does the survival and sudden death of quadripartite steering in curved spacetime truly depend on multi-directionality?},
  author={Liu, Xiaobao and Liu, Wentao and Shang, Si-Han and Wu, Shu-Min},
  journal={arXiv preprint arXiv:2511.01561},
  year={2025}
}

@article{wu2025can,
  title={Can Hawking effect of multipartite state protect quantum resources in Schwarzschild black hole?},
  author={Wu, Shu-Min and Teng, Xiao-Wei and Yang, Hui-Chen and Xu, Rui-Yang and Barros, PHM and Costa, HAS},
  journal={arXiv preprint arXiv:2509.15002},
  year={2025}
}

@article{li2025multiqubit,
  title={Multiqubit coherence of mixed states near event horizon},
  author={Li, Wen-Mei and Lu, Jianbo and Wu, Shu-Min},
  journal={arXiv preprint arXiv:2505.07476},
  year={2025}
}

@article{li2024bosonic,
  title={Bosonic and fermionic coherence of N-partite states in the background of a dilaton black hole},
  author={Li, Wen-Mei and Wu, Shu-Min},
  journal={J. High Energ. Phys.},
  volume={2024},
  number={9},
  pages={1--20},
  year={2024},
  publisher={Springer}
}

@article{wu2024does,
  title={Does anti-Unruh effect assist quantum entanglement and coherence?},
  author={Wu, Shu-Min and Teng, Xiao-Wei and Li, Jin-Xuan and Zeng, Hao-Sheng and Liu, Tonghua},
  journal={New J. Phys.},
  volume={26},
  number={4},
  pages={043016},
  year={2024},
  publisher={IOP Publishing}
}

@article{li2025entropic,
  title={Entropic uncertainty and coherence in Einstein-Gauss-Bonnet gravity},
  author={Li, Wen-Mei and Lu, Jianbo and Wu, Shu-Min},
  journal={arXiv preprint arXiv:2510.13167},
  year={2025}
}

@article{barros2025velocity,
  title={Velocity effects slightly mitigating the quantumness degradation of an Unruh-DeWitt detector},
  author={Barros, PHM and Wu, Shu-Min and Almeida, CAS and Costa, HAS},
  journal={arXiv preprint arXiv:2510.01280},
  year={2025}
}

@article{leggett1980macroscopic,
  author = {Leggett, A. J.},
    title = {Macroscopic Quantum Systems and the Quantum Theory of Measurement},
    journal = {Progress of Theoretical Physics Supplement},
    volume = {69},
    pages = {80-100},
    year = {1980},
    month = {03},
    issn = {0375-9687},
    doi = {10.1143/PTP.69.80},
    url = {https://doi.org/10.1143/PTP.69.80},
    eprint = {https://academic.oup.com/ptps/article-pdf/doi/10.1143/PTP.69.80/5356381/69-80.pdf},
}

@article{bibak2024quantum,
  title = {Quantum Coherence in Networks},
  author = {Bibak, Fatemeh and Del Santo, Flavio and Daki\ifmmode \acute{c}\else \'{c}\fi{}, Borivoje},
  journal = {Phys. Rev. Lett.},
  volume = {133},
  issue = {23},
  pages = {230201},
  numpages = {6},
  year = {2024},
  month = {Dec},
  publisher = {American Physical Society},
  doi = {10.1103/PhysRevLett.133.230201},
  url = {https://link.aps.org/doi/10.1103/PhysRevLett.133.230201}
}

@article{shi2022entanglement,
  title = {Entanglement, Coherence, and Extractable Work in Quantum Batteries},
  author = {Shi, Hai-Long and Ding, Shu and Wan, Qing-Kun and Wang, Xiao-Hui and Yang, Wen-Li},
  journal = {Phys. Rev. Lett.},
  volume = {129},
  issue = {13},
  pages = {130602},
  numpages = {6},
  year = {2022},
  month = {Sep},
  publisher = {American Physical Society},
  doi = {10.1103/PhysRevLett.129.130602},
  url = {https://link.aps.org/doi/10.1103/PhysRevLett.129.130602}
}

@article{ahnefeld2022coherence,
  title = {Coherence as a Resource for Shor's Algorithm},
  author = {Ahnefeld, Felix and Theurer, Thomas and Egloff, Dario and Matera, Juan Mauricio and Plenio, Martin B.},
  journal = {Phys. Rev. Lett.},
  volume = {129},
  issue = {12},
  pages = {120501},
  numpages = {7},
  year = {2022},
  month = {Sep},
  publisher = {American Physical Society},
  doi = {10.1103/PhysRevLett.129.120501},
  url = {https://link.aps.org/doi/10.1103/PhysRevLett.129.120501}
}

@article{karli2024controlling,
  author={Karli, Yusuf
and Vajner, Daniel A.
and Kappe, Florian
and Hagen, Paul C. A.
and Hansen, Lena M.
and Schwarz, Ren{\'e}
and Bracht, Thomas K.
and Schimpf, Christian
and Covre da Silva, Saimon F.
and Walther, Philip
and Rastelli, Armando
and Axt, Vollrath Martin
and Loredo, Juan C.
and Remesh, Vikas
and Heindel, Tobias
and Reiter, Doris E.
and Weihs, Gregor},
title={Controlling the photon number coherence of solid-state quantum light sources for quantum cryptography},
journal={npj Quantum Information},
year={2024},
month={Jan},
day={27},
volume={10},
number={1},
pages={17},
issn={2056-6387},
doi={10.1038/s41534-024-00811-2},
url={https://doi.org/10.1038/s41534-024-00811-2}
}

@article{yamauchi2024modulation,
  author={Yamauchi, Akio
and Fujiwara, Saiya
and Kimizuka, Nobuo
and Asada, Mizue
and Fujiwara, Motoyasu
and Nakamura, Toshikazu
and Pirillo, Jenny
and Hijikata, Yuh
and Yanai, Nobuhiro},
title={Modulation of triplet quantum coherence by guest-induced structural changes in a flexible metal-organic framework},
journal={Nature Communications},
year={2024},
month={Sep},
day={02},
volume={15},
number={1},
pages={7622},
issn={2041-1723},
doi={10.1038/s41467-024-51715-w},
url={https://doi.org/10.1038/s41467-024-51715-w}
}

@article{wang2024physical,
  author={Wang, Yue
and Hu, Yixiao
and Guo, Jian-Ping
and Gao, Jun
and Song, Bo
and Jiang, Lei},
title={A physical derivation of high-flux ion transport in biological channel via quantum ion coherence},
journal={Nature Communications},
year={2024},
month={Aug},
day={21},
volume={15},
number={1},
pages={7189},
issn={2041-1723},
doi={10.1038/s41467-024-51045-x},
url={https://doi.org/10.1038/s41467-024-51045-x}
}

@article{araujo2025non,
  title={A non-commutative Kalb-Ramond black hole},
  author={Ara{\'u}jo Filho, AA and Heidari, N and Lobo, Iarley P},
  journal={J. Cosmol. Astropart. Phys.},
  volume={2025},
  number={09},
  pages={076},
  year={2025},
  publisher={IOP Publishing}
}

@article{ARAUJOFILHO2025,
title = {Particle production induced by a Lorentzian non-commutative spacetime},
journal = {Ann. Phys.},
volume = {481},
pages = {170167},
year = {2025},
issn = {0003-4916},
doi = {https://doi.org/10.1016/j.aop.2025.170167},
url = {https://www.sciencedirect.com/science/article/pii/S0003491625002490},
author = {A. A. {Araújo Filho}}
}

@article{araujo2025does,
  title={How does non-metricity affect particle creation and evaporation in bumblebee gravity?},
  author={Ara{\'u}jo Filho, A. A.},
  journal={J. Cosmol. Astropart. Phys.},
  volume={2025},
  number={06},
  pages={026},
  year={2025},
  publisher={IOP Publishing}
}

@article{araujo2025particle,
  title={Particle creation and evaporation in Kalb-Ramond gravity},
  author={Ara{\'u}jo Filho, AA},
  journal={J. Cosmol. Astropart. Phys.},
  volume={2025},
  number={04},
  pages={076},
  year={2025},
  publisher={IOP Publishing}
}

@Article{mukherjee2022unruh,
author={Mukherjee, Arnab and Gangopadhyay, Sunandan and Majumdar, A. S.},
title={Unruh quantum Otto engine in the presence of a reflecting boundary},
journal={J. High Energ. Phys.},
year={2022},
month={Sep},
day={14},
volume={2022},
number={9},
pages={105},
issn={1029-8479},
doi={10.1007/JHEP09(2022)105},
url={https://doi.org/10.1007/JHEP09(2022)105}
}

@article{mukherjee2023fulling,
  title = {Fulling-Davies-Unruh effect for accelerated two-level single and entangled atomic systems},
  author = {Mukherjee, Arnab and Gangopadhyay, Sunandan and Majumdar, Archan S.},
  journal = {Phys. Rev. D},
  volume = {108},
  issue = {8},
  pages = {085018},
  numpages = {26},
  year = {2023},
  month = {Oct},
  publisher = {American Physical Society},
  doi = {10.1103/PhysRevD.108.085018},
  url = {https://link.aps.org/doi/10.1103/PhysRevD.108.085018}
}

@article{mukherjee2024quantum,
  title = {Quantum coherence measures for generalized Gaussian wave packets under a Lorentz boost},
  author = {Mukherjee, Arnab and Sen, Soham and Gangopadhyay, Sunandan},
  journal = {Phys. Rev. A},
  volume = {110},
  issue = {5},
  pages = {052413},
  numpages = {15},
  year = {2024},
  month = {Nov},
  publisher = {American Physical Society},
  doi = {10.1103/PhysRevA.110.052413},
  url = {https://link.aps.org/doi/10.1103/PhysRevA.110.052413}
}

@article{mukherjee2025quantum,
      author={Arnab Mukherjee and Soham Sen and Sunandan Gangopadhyay},
      title={Quantum coherence measures in entangled atomic systems}, 
      eprint={2511.20371},
      journal = {},
      archivePrefix={arXiv},
      primaryClass={quant-ph},
      month = {11},
      year={2025}
}

@article{mukherjee2024enhancement,
    author = {Mukherjee, Arnab and Gangopadhyay, Sunandan and Majumdar, A. S.},
    title = {{Enhancement of an Unruh-DeWitt battery performance through quadratic environmental coupling}},
    eprint = {2411.02849},
    journal = {},
    archivePrefix = {arXiv},
    primaryClass = {gr-qc},
    month = {11},
    year = {2024}
}

@article{mukherjee2025single,
  title = {{Single and Entangled Atomic Systems in Thermal Bath and the Fulling-Davies-Unruh Effect}},
  author = {Mukherjee, Arnab and Gangopadhyay, Sunandan and Majumdar, Archan S.},
  journal = {Quanta},
  volume = {14},
  number = {1},
  pages = {1--27},
  year = {2025},
  doi = {10.12743/quanta.v14i1.280},
  url = {https://doi.org/10.12743/quanta.v14i1.280}
}

@article{hummer2016renormalized,
  title = {Renormalized Unruh-DeWitt particle detector models for boson and fermion fields},
  author = {H\"ummer, Daniel and Mart\'{\i}n-Mart\'{\i}nez, Eduardo and Kempf, Achim},
  journal = {Phys. Rev. D},
  volume = {93},
  issue = {2},
  pages = {024019},
  numpages = {50},
  year = {2016},
  month = {Jan},
  publisher = {American Physical Society},
  doi = {10.1103/PhysRevD.93.024019},
  url = {https://link.aps.org/doi/10.1103/PhysRevD.93.024019}
}

@article{sachs2017entanglement,
  title = {Entanglement harvesting and divergences in quadratic Unruh-DeWitt detector pairs},
  author = {Sachs, Allison and Mann, Robert B. and Mart\'{\i}n-Mart\'{\i}nez, Eduardo},
  journal = {Phys. Rev. D},
  volume = {96},
  issue = {8},
  pages = {085012},
  numpages = {17},
  year = {2017},
  month = {Oct},
  publisher = {American Physical Society},
  doi = {10.1103/PhysRevD.96.085012},
  url = {https://link.aps.org/doi/10.1103/PhysRevD.96.085012}
}

@Article{wu2023birth,
author={Wu, Dawei and Tang, Shan-Chang and Shi, Yu},
title={Birth and death of entanglement between two accelerating Unruh-DeWitt detectors coupled with a scalar field},
journal={J. High Energ. Phys.},
year={2023},
month={Dec},
day={05},
volume={2023},
number={12},
pages={37},
issn={1029-8479},
doi={10.1007/JHEP12(2023)037},
url={https://doi.org/10.1007/JHEP12(2023)037}
}

@article{ARAUJOFILHO2025117174,
title = {Axisymmetric black hole in a non–commutative gauge theory: Classical and quantum gravity effects},
journal = {Nucl. Phys. B},
volume = {1020},
pages = {117174},
year = {2025},
issn = {0550-3213},
doi = {https://doi.org/10.1016/j.nuclphysb.2025.117174},
url = {https://www.sciencedirect.com/science/article/pii/S0550321325003839},
author = {A.A. {Araújo Filho} and N. Heidari and A. Övgün}
}

@article{Heidari:2025iiv,
    author = "Heidari, N. and Ara{\'u}jo Filho, A. A. and Lobo, Iarley P.",
    title = "{Non-commutativity in Hayward spacetime}",
    eprint = "2503.17789",
    archivePrefix = "arXiv",
    primaryClass = "gr-qc",
    doi = "10.1088/1475-7516/2025/09/051",
    journal = "J. Cosmol. Astropart. Phys.",
    volume = "09",
    pages = "51",
    year = "2025"
}

@article{AraujoFilho:2025jcu,
    author = "Ara{\'u}jo Filho, A. A. and Heidari, N. and Lobo, Iarley P.",
    title = "{A non-commutative Kalb-Ramond black hole}",
    eprint = "2507.17390",
    archivePrefix = "arXiv",
    primaryClass = "gr-qc",
    doi = "10.1088/1475-7516/2025/09/076",
    journal = "J. Cosmol. Astropart. Phys.",
    volume = "09",
    pages = "76",
    year = "2025"
}

\end{document}